# Measurement of the $WW + WZ$ cross section and limits on anomalous triple gauge couplings using final states with one lepton, missing transverse momentum, and two jets with the ATLAS detector at $\sqrt{s} = 7$ TeV

The ATLAS Collaboration

**Abstract**

The production of a $W$ boson decaying to $e\nu$ or $\mu\nu$ in association with a $W$ or $Z$ boson decaying to two jets is studied using $4.6\ \mathrm{fb}^{-1}$ of proton–proton collision data at $\sqrt{s} = 7$ TeV recorded with the ATLAS detector at the LHC. The combined $WW+WZ$ cross section is measured with a significance of $3.4\sigma$ and is found to be $68 \pm 7\ (\mathrm{stat.}) \pm 19\ (\mathrm{syst.})\ \mathrm{pb}$, in agreement with the Standard Model expectation of $61.1 \pm 2.2\ \mathrm{pb}$. The distribution of the transverse momentum of the dijet system is used to set limits on anomalous contributions to the triple gauge coupling vertices and on parameters of an effective-field-theory model.

# Measurement of the $WW + WZ$ cross section and limits on anomalous triple gauge couplings using final states with one lepton, missing transverse momentum, and two jets with the ATLAS detector at $\sqrt{s} = 7$ TeV

**The ATLAS Collaboration**

Abstract:

The production of a $W$ boson decaying to $e\nu$ or $\mu\nu$ in association with a $W$ or $Z$ boson decaying to two jets is studied using 4.6 fb$^{-1}$ of proton–proton collision data at $\sqrt{s} = 7$ TeV recorded with the ATLAS detector at the LHC. The combined $WW + WZ$ cross section is measured with a significance of $3.4\sigma$ and is found to be $68 \pm 7$ (stat.) $\pm\, 19$ (syst.) $pb$, in agreement with the Standard Model expectation of $61.1 \pm 2.2$ pb. The distribution of the transverse momentum of the dijet system is used to set limits on anomalous contributions to the triple gauge coupling vertices and on parameters of an effective-field-theory model.

## Contents

## 1 Introduction

The study of vector boson pair production at the Large Hadron Collider (LHC) provides an important test of the electroweak sector of the Standard Model (SM) at the highest available energies. Deviations observed in the total or differential cross sections from the SM predictions may arise from anomalous triple gauge boson interactions [1] or from new particles decaying into vector bosons [2]. Vector boson pair production is also an important source of background in studies of the Higgs boson and in searches for signals of physics beyond the SM.

The cross sections for $WW$ and $WZ$ production at the LHC have previously been measured in fully leptonic final states [3–6]. The semileptonic final states suffer from larger backgrounds from $W$ or $Z$ boson production in association with jets, but benefit from significantly larger branching fractions than the fully leptonic states and thus represent important complementary measurements. In this paper the $WW + WZ$ cross section is measured in the $\ell\nu$jj ($\ell = e, \mu$) final state using a data sample of proton–proton ($pp$) collisions with an integrated luminosity of 4.6 fb$^{-1}$ collected by the ATLAS detector at the LHC. In addition, the reconstructed dijet transverse momentum distribution is used to set limits on anomalous contributions to the triple gauge coupling vertices (aTGCs), after requiring that the dijet invariant mass is close to the mass of the $W$ or $Z$ boson.

The combined $WW+WZ$ production cross section (hereafter, $WV$ cross section, where $V = W, Z$) has been measured in the $\ell\nu$jj final state in proton–antiproton collisions at the Tevatron collider by both the CDF [7] and D0 [8] collaborations, and more recently in $pp$ collisions by the CMS [9] collaboration. Limits on anomalous triple gauge couplings in $WV \to \ell\nu$jj production have also been presented by CDF [10], D0 [11], and CMS [9].

This paper is organised as follows. The overall analysis strategy is described in section 2 and a short description of the ATLAS detector is given in section 3. The Monte Carlo (MC) simulation used for the signal and background modelling is summarized in section 4. Details of the object and event reconstruction and of the event selection are given in sections 5 and 6 respectively. The method to estimate the signal and background processes is discussed in section 7. The cross section measurement is detailed in section 8 and the systematic uncertainties are described in section 9. The results of the cross-section measurement are summarized in section 10, and the extraction of the anomalous triple gauge coupling limits is discussed in section 11. Finally, conclusions are drawn in section 12.

## 2 Analysis strategy

Candidate $WV \to \ell\nu$jj events are required to contain exactly one lepton (electron or muon), large missing transverse momentum $E_{\mathrm{T}}^{\mathrm{miss}}$, and exactly two jets. The selected events are accepted if they pass a set of kinematic cuts chosen to enhance the signal-to-background ratio. The invariant mass distribution of the two jets ($m_{\mathrm{jj}}$), representing the candidate decay products of the hadronically decaying boson, is obtained from all the selected events. The $WW + WZ$ signal yield ($N^{WV}$) is obtained by performing a binned maximum-likelihood fit to the $m_{\mathrm{jj}}$ distribution using templates based on MC simulations. The fit is performed on events in an $m_{\mathrm{jj}}$ range much larger than the range where the signal peaks, allowing the nearly signal-free $m_{\mathrm{jj}}$ regions to constrain the rate of the $W +$jets events, which are the largest background. Because of the finite dijet mass resolution, there is considerable overlap between the $m_{\mathrm{jj}}$ peaks from $WW \to \ell\nu q\bar{q}'$ and $WZ \to \ell\nu q\bar{q}$ decays. Given the expected uncertainties in this measurement, and the relatively small contribution from the $WZ$ process (about 20% of the total signal yield), no attempt is made to distinguish between the $WW$ and $WZ$ contributions in this analysis. Instead, the signal yield is obtained under the assumption that the ratio of the $WW$ and $WZ$ cross sections is equal to the SM prediction.

The fiducial cross section ($\sigma_{\mathrm{fid}}$) is evaluated from the measured signal yield. The fiducial phase space is defined to be as close as possible to the phase space defined by the reconstructed event selection. The fiducial cross-section measurement is obtained as:

$$\sigma_{\mathrm{fid}} = \sum_{\ell=e,\mu} \frac{N_{\ell}^{WV}}{\mathcal{L} \cdot D_{\mathrm{fid},\ell}}\,, \qquad (2.1)$$

where $\mathcal{L}$ is the integrated luminosity and $D_{\text{fid},\ell}$ are factors that correct for the difference between the number of $WV \to \ell\nu\text{jj}$ events produced in the fiducial phase space and the number of reconstructed events passing the event selection. The total cross section ($\sigma_{\text{tot}}$) is obtained by extrapolating the fiducial cross section to the full phase space using theoretical predictions:

$$\sigma_{\text{tot}} = \sum_{\ell=e,\mu} \frac{N_\ell^{WV}}{\mathcal{L} \cdot D_{\text{tot},\ell}}, \tag{2.2}$$

where $D_{\text{tot},\ell}$ are factors that depend on acceptances, reconstruction efficiencies, and the branching fractions for $WW \to \ell\nu\text{jj}$ and $WZ \to \ell\nu\text{jj}$. Details of the maximum-likelihood fit and the precise definition of the fiducial-volume and of the factors $D_{\text{fid},\ell}$ and $D_{\text{tot},\ell}$ are given in section 8.

Lastly, the transverse momentum distribution of the hadronically decaying $V$ candidates ($p_{\text{T}jj}$) is used to set limits on the aTGCs affecting the $WWZ$ and $WW\gamma$ vertices. The event selection is the same as the one used for the cross-section measurement, except that the dijet mass is required to be close to the masses of the $W/Z$ bosons in order to increase the signal-to-background ratio. The aTGC limits are calculated by performing a binned maximum-likelihood fit to the $p_{\text{T}jj}$ distributions. The ratio of the $WW$ and $WZ$ cross sections at each aTGC point is assumed to be that predicted by theory, including the aTGC contribution. Details of the limit extraction are given in section 11.

## 3 The ATLAS detector

The ATLAS detector [12] is a general-purpose particle detector with cylindrical geometry[1] which consists of several sub-detectors surrounding the interaction point, and covering almost the full solid angle. The trajectories and momenta of charged particles are measured within the pseudorapidity region $|\eta| < 2.5$ by multi-layer silicon pixel and microstrip detectors and a transition radiation tracker. The tracking system is located in a superconducting solenoid producing a 2 T magnetic field and is surrounded by a high-granularity liquid-argon (LAr) sampling electromagnetic (EM) calorimeter with coverage up to $|\eta| = 3.2$. The EM calorimeter is split into a barrel section ($|\eta| < 1.475$) and endcaps ($1.375 < |\eta| < 3.2$). A scintillating tile hadronic calorimeter using steel as absorber provides coverage in the range $|\eta| < 1.7$. In the forward region, LAr calorimeters provide electromagnetic and hadronic measurements and extend the coverage to $|\eta| < 4.9$. The muon spectrometer surrounds the ATLAS calorimeter system and it operates in a toroidal magnetic field provided by air-core superconducting magnets and includes tracking chambers for precise muon momentum measurement up to $|\eta| = 2.7$ and trigger chambers covering the range $|\eta| < 2.4$. The online event selection is based on a three-level trigger system. The hardware-based Level-1 trigger uses a subset of the detector data to reduce the event rate from 20 MHz to below 75 kHz. Two subsequent software-level triggers further reduce the rate to about 300 Hz using the complete detector information.

[1] ATLAS uses a right-handed coordinate system with its origin at the nominal interaction point (IP) in the centre of the detector and the $z$-axis coinciding with the axis of the beam pipe. The $x$-axis points from the interaction point to the centre of the LHC ring, and the $y$-axis points upward. Cylindrical coordinates $(r, \phi)$ are used in the transverse plane, $\phi$ being the azimuthal angle around the beam pipe. The pseudorapidity $\eta$ is defined in terms of the polar angle $\theta$ as $\eta = -\ln[\tan(\theta/2)]$.

## 4 Simulated event samples

Simulated event samples are used to model both the signal and all background processes except for the multijet background, which is estimated using a data-driven procedure. The signal and background MC samples are processed using the ATLAS detector simulation [13] based on GEANT4 [14] and the same reconstruction algorithms as used for collision data. The simulation includes the modelling of additional $pp$ interactions in the same and neighbouring bunch crossings (pile-up).

Diboson signal events are generated using MC@NLO v4.07 [15] interfaced to HERWIG [16, 17] for the parton showering and hadronisation and to JIMMY[2] [18] for the modelling of the underlying event. The on-shell gauge bosons are generated in MC@NLO and are subsequently decayed by HERWIG. This leads to a zero width for the decayed $W/Z$ bosons and to the loss of the spin-correlation information for the decay products. The effects arising from this generation procedure are studied and considered, where needed, as systematic uncertainties. The CT10 [19] parton distribution function (PDF) set is used. The diboson samples are normalised to the next-to-leading-order (NLO) cross sections of $43.7 \pm 1.9$ pb and $17.4 \pm 1.1$ pb for $WW$ and $WZ$, respectively. The central values of the diboson cross sections are estimated using MC@NLO, with factorisation and renormalisation scales equal to $\sqrt{(m_W^2 + p_{\mathrm{T},W}^2 + m_V^2 + p_{\mathrm{T},V}^2)/2}$. The uncertainties are evaluated by varying the scales, the PDF, and $\alpha_\mathrm{s}$. The combined PDF+$\alpha_\mathrm{s}$ uncertainties are estimated by varying them within their 68% confidence-level (CL) limits, following the procedure in ref. [20]. The $gg \to WW$ and $H \to WW$ processes are not included in the signal samples nor in the cross-section prediction, since their contributions are small compared to the expected sensitivity of this measurement. The $gg \to WW$ process would increase the total predicted $WV$ cross section by about 2–4%. The $H \to WW$ process would increase the $WV$ cross section by about 5%, but after applying all event selection criteria (see section 6), it would only increase the expected number of signal events by about 2%. The $\gamma\gamma \to WW$ process [21] is also neglected. While this paper was in preparation, a next-to-next-to-leading-order (NNLO) calculation of $W^+W^-$ production was published [22]. The NNLO corrections would increase the total $WV$ cross section by about 4% beyond the $gg \to WW$ and $H \to WW$ corrections already mentioned. Additional signal samples generated with PYTHIA[3] [23] are used for systematic studies.

The dominant background to the $WV \to \ell\nu\mathrm{jj}$ process is vector boson $W/Z$ production in association with jets, which is modelled using ALPGEN v2.13 [24] with CTEQ6L1 [25] for the PDF, interfaced to HERWIG and JIMMY. The $W/Z$ + jets cross sections predicted by ALPGEN are scaled to the QCD NNLO inclusive cross section [26] times branching fraction for a single lepton species: $\sigma(W \to \ell\nu) = 10.46 \pm 0.42$ nb and $\sigma(Z/\gamma^* \to \ell\ell) =$

[2]The versions HERWIG v6.520 and JIMMY v4.31 are used whenever these generators are mentioned.
[3]Whenever PYTHIA is mentioned, the version 6.425 is used.

$1.070 \pm 0.054$ nb for invariant masses of the two leptons ($m_{\ell\ell}$) $> 40$ GeV. Production of a $W$ or $Z$ boson plus heavy-flavour jets is also modeled using the ALPGEN+HERWIG+JIMMY generator combination described above, and overlap with the inclusive $W/Z+$jets samples is removed to avoid double-counting. Samples generated using SHERPA v1.4.1 [27–30] with CT10 PDFs are used for cross-checks.

Samples of $t\bar{t}$ events are produced using MC@NLO v4.01 [31] with the CT10 PDF set, interfaced to HERWIG and JIMMY. The $t\bar{t}$ cross section is $\sigma_{t\bar{t}} = 177^{+10}_{-11}$ pb for a top quark mass of 172.5 GeV. It has been calculated at NNLO in QCD including resummation of next-to-next-to-leading logarithmic (NNLL) soft gluon terms with TOP++2.0 [32–37]. Samples of $t\bar{t}$ events generated with ACERMC v3.8 [38] interfaced to PYTHIA are also considered for systematic uncertainty studies.

Single-top events from the $Wt$ and s-channel processes are generated using MC@NLO v4.01 [39, 40] interfaced to HERWIG and JIMMY with cross sections of $15.7 \pm 1.2$ pb [41] and $4.6 \pm 0.2$ pb [42], respectively. The CT10 PDF set is used. The single-top t-channel process is generated using ACERMC v3.8 + PYTHIA with the MRST LO** [43] PDF set, using a cross section of $64.6^{+2.6}_{-1.7}$ pb [44].

The $ZZ$ diboson background process is generated using HERWIG with MRST LO** PDFs. It is normalised to the NLO cross section of $5.96 \pm 0.3$ pb ($m_{\ell\ell} > 60$ GeV), estimated with MCFM [45]. The uncertainty is evaluated using the same procedure as for the diboson signal. The $W\gamma$ process is generated with MADGRAPH v4 [46] interfaced to PYTHIA. After the selection criteria are applied, the contribution of this process to the background is very small (less than 0.5% of the total), and so it is neglected.

## 5 Object and event reconstruction

Events were selected by a single-lepton (electron or muon) trigger with a threshold on the transverse energy ($E_{\mathrm{T}}$) in the electron case or on the transverse momentum ($p_{\mathrm{T}}$) in the muon case. The $p_{\mathrm{T}}$ threshold for the single-muon trigger was 18 GeV, while for electrons it was required to be $E_{\mathrm{T}} > 20$ GeV for the early part of data-taking and $E_{\mathrm{T}} > 22$ GeV when the instantaneous luminosity of the LHC increased.

Proton–proton collision events are identified by requiring that the events have at least one reconstructed vertex with at least three associated tracks with transverse momentum $p_{\mathrm{T,track}} > 0.4$ GeV. If two or more such vertices are found, the one with the largest sum of $p^2_{\mathrm{T,track}}$ is considered to be the primary vertex.

Electron candidates are formed by associating clusters of cells in the EM calorimeter with tracks reconstructed in the inner detector [47]. The transverse energy ($E_{\mathrm{T}}$), calculated from the cluster energy and the track direction, must be greater than 25 GeV, in order to be in the region with maximum trigger efficiency. Candidates are accepted if they lie in the region $|\eta| < 2.47$, excluding the transition region between the barrel and endcap EM calorimeters, $1.37 < |\eta| < 1.52$. The candidate must satisfy "tight" identification criteria described in ref. [47]. For the electron-candidate track, the ratio of the transverse impact parameter, $d_0$, to its uncertainty, $\sigma(d_0)$, must satisfy $|d_0/\sigma(d_0)| < 10$. The longitudinal impact parameter, $z_0$, must have an absolute value less than 1 mm. Both $d_0$ and $z_0$ are measured with respect to the primary vertex. To ensure isolation from surrounding particles, calorimetric and tracking criteria are applied. The total calorimeter $E_{\mathrm{T}}$ in a cone of size $\Delta R = \sqrt{\Delta\phi^2 + \Delta\eta^2} = 0.3$ around the electron candidate, excluding any $E_{\mathrm{T}}$ associated with the candidate itself, must be less than 14% of the electron $E_{\mathrm{T}}$ value. The calorimeter response is corrected for the additional energy deposited by pile-up. In addition, the scalar sum of the $p_{\mathrm{T}}$ of the tracks within $\Delta R = 0.3$ of the electron candidate (not including the electron track) must be less than 13% of the electron $p_{\mathrm{T}}$ value.

Muon candidates are identified [48] by associating tracks reconstructed in the muon spectrometer with tracks reconstructed in the inner detector. The momentum of the combined muon track is calculated from the momenta of the two tracks, correcting for the energy loss in the calorimeter. Muon candidates must satisfy $p_{\mathrm{T}} > 25$ GeV and $|\eta| < 2.4$. The $p_{\mathrm{T}}$ threshold is chosen to be well within the plateau of the trigger efficiency. Muon candidates must also be consistent with originating from the primary vertex, in order to reject muons from cosmic-ray interactions and to reduce background from heavy-flavour decays. Specifically, the $d_0$ significance must satisfy $|d_0/\sigma(d_0)| < 3$ and $|z_0|$ must be less than 1 mm. To reduce misidentification and improve the muon momentum resolution, requirements on the minimum number of hits in the various detectors are applied to the muon tracks. Isolated muons are selected with a requirement that the scalar sum of the $p_{\mathrm{T}}$ of the tracks within $\Delta R = 0.3$ of the muon (not including the muon track) be less than 15% of the muon $p_{\mathrm{T}}$, and that the total calorimeter $E_{\mathrm{T}}$ in a cone of $\Delta R = 0.3$ around the muon candidate (excluding $E_{\mathrm{T}}$ associated with the muon) be less than 14% of the muon $p_{\mathrm{T}}$. The electron and muon isolation requirements are the same as used in ref. [3].

Corrections are applied to MC events in order to account for differences between data and MC simulation in the trigger and identification efficiencies, and in the lepton momentum and energy scale and resolution. The trigger and reconstruction efficiency scale factors are measured using the tag-and-probe method on events with $Z$-boson candidates events [47, 48]. The lepton momenta are calibrated with scale factors obtained by comparing the reconstructed mass distribution of $Z$ boson candidate events in data with that of simulated events [48, 49].

Jets are reconstructed from calorimeter energy clusters by using an anti-$k_t$ algorithm [50, 51] with a radius parameter of 0.4. The selected jets must satisfy $E_{\mathrm{T}} > 25$ GeV and $|\eta| < 2.8$. Reconstructed jets are corrected for the non-compensating calorimeter response, upstream material and other effects using $p_{\mathrm{T}}$- and $\eta$-dependent correction factors derived from MC and validated with test-beam and collision data [51]. Jets consistent with being produced from pile-up interactions are identified using the Jet Vertex Fraction variable (JVF). This variable is calculated using tracks that are associated with the jet, and is defined as the ratio of the scalar $p_{\mathrm{T}}$ sum of the associated tracks that originate from the primary vertex to the scalar $p_{\mathrm{T}}$ sum of all associated tracks. Jets that are within $|\eta| < 2.5$ are retained if they have JVF larger than 75% or if they have no associated track. The efficiency of this cut is $\sim 95\%$ up to $|\eta| < 2.5$ and is well modelled by the MC simulation. Jets are required to satisfy quality criteria and to lie at a distance $\Delta R > 0.5$ from well-identified leptons.

The $E_{\mathrm{T}}^{\mathrm{miss}}$ is estimated from reconstructed electrons with $|\eta| < 2.47$, muons with $|\eta| < 2.7$, jets with $|\eta| < 4.9$, and clusters of energy in the calorimeter not associated with

reconstructed objects having $|\eta| < 4.5$ [52]. The energy clusters are calibrated to the EM scale or the hadronic energy scale according to cluster characteristics. The expected energy deposit of identified muons in the calorimeter is subtracted.

## 6 Event selection

The $WV$ candidates are selected by requiring exactly one high-$p_{\mathrm{T}}$ lepton, missing transverse momentum, and exactly two jets. Events are required to contain exactly one reconstructed lepton candidate with $p_{\mathrm{T}} > 25$ GeV; events with more than one identified lepton are rejected in order to suppress the $Z$+jets and $t\bar{t}$ backgrounds. The lepton candidate must be the one that triggered the event. Furthermore, events are required to have $E_{\mathrm{T}}^{\mathrm{miss}} > 30$ GeV in order to account for the presence of the unobserved neutrino from the $W \to \ell\nu$ decays. The transverse mass of the leptonically decaying $W$ boson candidate is defined as $m_{\mathrm{T}} = \sqrt{2 \cdot p_{\mathrm{T}}(\ell) \cdot E_{\mathrm{T}}^{\mathrm{miss}} \cdot (1 - \cos\Delta\phi)}$, where $\Delta\phi$ is the azimuthal angle between the lepton momentum and missing transverse momentum vectors, and is required to satisfy $m_{\mathrm{T}} > 40$ GeV. The $E_{\mathrm{T}}^{\mathrm{miss}}$ and $m_{\mathrm{T}}$ criteria highly suppress the multijet background. To further suppress the multijet background the azimuthal angular separation between the leading jet transverse momentum and the missing transverse momentum vectors must fulfil $|\Delta\phi(E_{\mathrm{T}}^{\mathrm{miss}}, j_1)| > 0.8$. Backgrounds containing top quark decays are highly reduced by vetoing events that contain more than two jets with $p_{\mathrm{T}} > 25$ GeV and $|\eta| < 2.8$. Events are required to contain exactly two jets with $|\eta| < 2.0$ and $p_{\mathrm{T}} > 25$ GeV, with a $p_{\mathrm{T}} > 30$ GeV requirement for the leading jet. In order to improve the signal-to-background ratio, the two jets are required to satisfy $|\Delta\eta(j_1, j_2)| < 1.5$. The angular distance between the two jets must satisfy $\Delta R(j_1, j_2) > 0.7$ if the $p_{\mathrm{T}}$ of the dijet system is less than 250 GeV. Finally, the dijet invariant mass must be in the range $25 < m_{\mathrm{jj}} < 250$ GeV. The selection criteria were optimised to both increase the signal-to-background ratio and select a phase space region well described by the Monte Carlo simulation. After applying all event selection criteria, 127 650 events are found in the electron channel and 134 846 in the muon channel.

## 7 Signal and background estimation

The shapes of the expected $m_{\mathrm{jj}}$ and $p_{\mathrm{T}jj}$ distributions are used as templates for the cross-section fit and for the aTGC limit calculation, respectively. The expected shapes and rates of the distributions for the $W$+jets, $Z$+jets, $t\bar{t}$, single-top, and signal processes are obtained from the MC simulation samples. The $W+$jets and $Z+$jets predicted rates are corrected using scale factors obtained with a data-driven method as explained below.

Multijet background events can pass the event selection if one of the jets is reconstructed as a lepton. The rate and shape of the multijet background are estimated with data-driven methods since the MC simulation does not reliably predict the rate of jets passing the lepton identification.

The data-driven method consists of two steps: the first one is designed to estimate the $m_{\mathrm{jj}}$, $p_{\mathrm{T}jj}$, and $E_{\mathrm{T}}^{\mathrm{miss}}$ shapes of the multijet background and the second one to measure its rate.

The first step exploits suitably modified lepton identification criteria to define data samples enriched in multijet background and with kinematic characteristics as close as possible to those of the standard selection. The lepton identification criteria are modified differently for the muon and electron channels.

For the muon channel, the multijet-enriched sample is obtained by applying the full selection but inverting the transverse impact parameter requirement ($|d_0/\sigma(d_0)| > 3$). The selected sample is composed of muons that do not originate from the primary vertex, as expected for muons produced from heavy-flavour decays in jets. For the electron channel, the multijet-enriched sample is obtained by applying the full nominal selection but requiring the electron candidate to satisfy the "medium" [53] identification criteria but not the "tight" ones. This results in a sample enriched in events with a jet that mimics an electron. Finally, the shape of the multijet background is obtained from the data in these multijet-enriched samples, after subtracting the MC-based prediction for non-multijet processes.

The second step uses the $E_{\mathrm{T}}^{\mathrm{miss}}$ shape of the multijet background, determined in the previous step, to obtain the multijet rate and a correction to the $W/Z+$jets normalisation. This is done fitting the $E_{\mathrm{T}}^{\mathrm{miss}}$ spectra obtained with the nominal selection but with the $E_{\mathrm{T}}^{\mathrm{miss}}$ requirement removed. The fit, performed in the range $0 < E_{\mathrm{T}}^{\mathrm{miss}} < 400$ GeV, extracts separate scale factors used to normalise the multijet and $W/Z+$jets samples. From this fit, the multijet contribution is extrapolated to the signal region ($E_{\mathrm{T}}^{\mathrm{miss}} > 30$ GeV) and is found to represent 5.3% and 3.7% of the events for the electron and muon channels respectively. The $W/Z+$jets scale factors obtained from this fit are close to one and well within the systematic uncertainty of the theoretical prediction both for the electron and muon channel.

Table 1 shows the expected number of events for the signal and for each background process after the full selection is applied. The numbers of events observed in data are also listed. The signal-to-background ratio in the subrange $60 < m_{\mathrm{jj}} < 120$ GeV is about 2%. Figure 1 shows the $m_{\mathrm{jj}}$ distributions for data and the SM prediction for the electron and muon channels prior to performing the maximum-likelihood fit to extract the signal $WV$ yield. The bottom plots in figure 1 show the ratios of data to the SM predictions overlaid with systematic uncertainty bands. The sources of systematic uncertainties and the strategy to evaluate them are discussed in section 9. The data distributions are well within the systematic uncertainty bands for all values of dijet mass for both channels.

## 8 Cross-section definition and fit method

As discussed in section 2, $WV \to \ell\nu\mathrm{jj}$ candidates are selected in a fiducial phase space designed to increase the signal-to-background ratio. The fiducial phase space, which is identical for the electron and muon channels, is defined for Monte Carlo events by applying to the particle-level objects a selection as close as possible to the analysis selection described in section 6. This selection requires a $W$ boson decaying leptonically and a $W$ or $Z$ boson decaying hadronically. $W \to \tau\nu$ decays are not included in the definition of the fiducial cross section.

The leptonically decaying $W$ boson is required to decay to an electron or a muon with $p_{\mathrm{T}} > 25$ GeV and $|\eta| < 2.47$. The lepton $p_{\mathrm{T}}$ is obtained by summing together the lepton

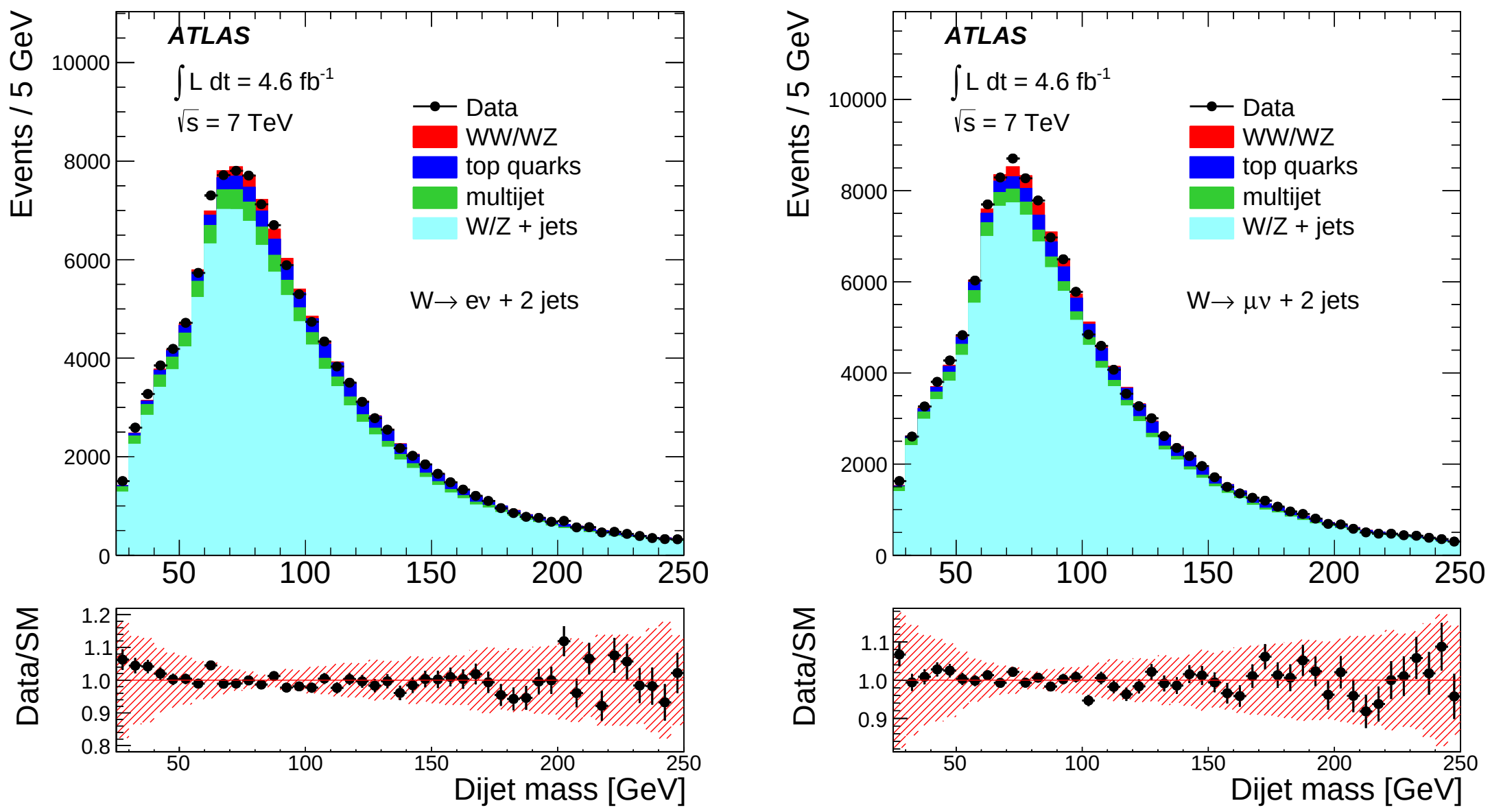

(a) (b)

**Figure 1**. Distributions of the dijet invariant mass for (a) the electron and (b) the muon channels before the likelihood fit. The error bars represent statistical uncertainties, and the stacked histograms are SM predictions. The lower panel displays the ratio of the data to the MC expectation. The systematic band contains only systematic uncertainties that affect the shape of the background and signal processes.

transverse momentum and the transverse momenta of all photons within $\Delta R = 0.1$ of the selected lepton. The transverse mass of the leptonically decayed $W$ boson is required to be $m_{\mathrm{T}} > 40$ GeV.

Events must contain a hadronically decaying $W$ or $Z$ boson and two particle-level jets separated by $\Delta R > 0.5$ from the selected leptons. Particle-level jets are reconstructed from particles with a mean decay length $c\tau > 10$ mm using the anti-$k_t$ algorithm with radius parameter $R = 0.4$. Decay products from leptonically decaying $W/Z$ bosons (including photons within $\Delta R = 0.1$ of the charged leptons) are excluded from the particle-level jets. The two selected jets must lie within $|\eta| < 2.0$ and have $p_{\mathrm{T}} > 25$ GeV with at least one of them having $p_{\mathrm{T}} > 30$ GeV. Events containing more than two particle-level jets with $p_{\mathrm{T}} > 25$ GeV and $|\eta| < 2.8$ are rejected. Moreover, the two selected jets must satisfy $|\Delta\eta(j_1, j_2)| < 1.5$, $25 < m_{\mathrm{jj}} < 250$ GeV and $\Delta R(j_1, j_2) > 0.7$.

The last condition is applied only if the transverse momentum of the dijet system is $p_{\mathrm{T}jj} < 250$ GeV. Finally the $E_{\mathrm{T}}^{\mathrm{miss}}$, defined as the transverse momentum of the neutrino from the leptonically decaying $W$ boson, is required to satisfy $E_{\mathrm{T}}^{\mathrm{miss}} > 30$ GeV and $|\Delta\phi(E_{\mathrm{T}}^{\mathrm{miss}}, j_1)| > 0.8$.

The signal event yield in the fiducial-volume is determined from a simultaneous maximum-likelihood fit to the $m_{\mathrm{jj}}$ distributions in the electron and muon channels. This method takes advantage of the difference between the shapes of the $m_{\mathrm{jj}}$ distributions of the various processes to separate the signal from the large underlying background. The $m_{\mathrm{jj}}$ templates, normalised to unit area, for the various processes contributing to the total expected $m_{\mathrm{jj}}$ distribution are shown in figure 2.

**Table 1**. Total number of events in data and expected yields for each process in the $e$ and $\mu$ channel. The multijet and $W/Z$+jets yields are obtained from the fit to the $E_{\mathrm{T}}^{\mathrm{miss}}$ distribution as explained in section 7. Uncertainties for the expected signal yields are based on the corresponding cross-section uncertainties, while for multijet and the other backgrounds the uncertainties correspond to the total rate uncertainty.

| Signal processes | $e$ | $\mu$ |
|---|---|---|
| $WW$ | 1435 ± 70 | 1603 ± 79 |
| $WZ$ | 334 ± 23 | 370 ± 26 |
| Background processes | | |
| $W$+ jets | $(107 \pm 21) \times 10^3$ | $(116 \pm 23) \times 10^3$ |
| $Z$+ jets | $(55 \pm 11)\times 10^2$ | $(46.3 \pm 9.3)\times 10^2$ |
| $t\bar{t}$ | $(47.2 \pm 7.1)\times 10^2$ | $(47.2 \pm 7.1)\times 10^2$ |
| Single-top | $(20.2 \pm 3.0)\times 10^2$ | $(20.5 \pm 3.1)\times 10^2$ |
| Multijet | $(67 \pm 10)\times 10^2$ | $(50.5 \pm 7.6)\times 10^2$ |
| $ZZ$ | 19.2 ± 3.8 | 21.1 ± 4.2 |
| Total SM prediction | $(128 \pm 17)\times 10^3$ | $(135 \pm 19)\times 10^3$ |
| Total Data | 127 650 | 134 846 |

Systematic uncertainties (described in section 9) on the signal and background normalisation as well as on the $m_{\mathrm{jj}}$ shapes are included by introducing nuisance parameters ($\vec{\alpha}$) into the fit. The combined likelihood function ($L$) is expressed as:

$$L(\beta,\ \vec{\alpha}) = \prod_{\ell}\prod_{b} \mathrm{Poisson}(n_{\ell b}|(\nu_{\ell b}^{\mathrm{bkg}} + \beta\nu_{\ell b}^{\mathrm{sig}})(\vec{\alpha})) \cdot \prod_{p} f_p(\alpha_p), \tag{8.1}$$

where $\beta$ is the parameter of interest extracted from the fit and is a multiplicative factor applied to the signal normalisation; $n_{\ell b}$ is the number of data events in bin $b$ and channel $\ell$, with $\ell = e, \mu$; $\nu_{\ell b}^{\mathrm{bkg}}$ and $\nu_{\ell b}^{\mathrm{sig}}$ are the number of expected events for background and signal processes respectively in bin $b$ and channel $\ell$; and $f_p$ are Gaussian constraints on the nuisance parameters $\alpha_p$. The expected number of signal events $\nu_{\ell b}^{\mathrm{sig}}$ contains contributions from both the $WW$ and $WZ$ processes. The measured signal yield $N_{\ell}^{WV}$ is obtained from the product of the fitted $\beta$ value and the expected number of signal events as $N_{\ell}^{WV} = \beta \cdot \sum_b \nu_{\ell b}^{\mathrm{sig}}$.

The diboson fiducial cross section ($\sigma_{\mathrm{fid}}$) is extracted from $N_{\ell}^{WV}$ using eq. (2.1). The factors $D_{\mathrm{fid},\ell}$ account for the fact that two processes, $WW \to \ell\nu\mathrm{jj}$ and $WZ \to \ell\nu\mathrm{jj}$, contribute to the signal yield with different cross sections, acceptances and correction factors and are defined as:

$$D_{\mathrm{fid},\ell} = f_{\mathrm{fid}}^{WW} \cdot C_{\ell}^{WW} + (1 - f_{\mathrm{fid}}^{WW}) \cdot C_{\ell}^{WZ}, \tag{8.2}$$

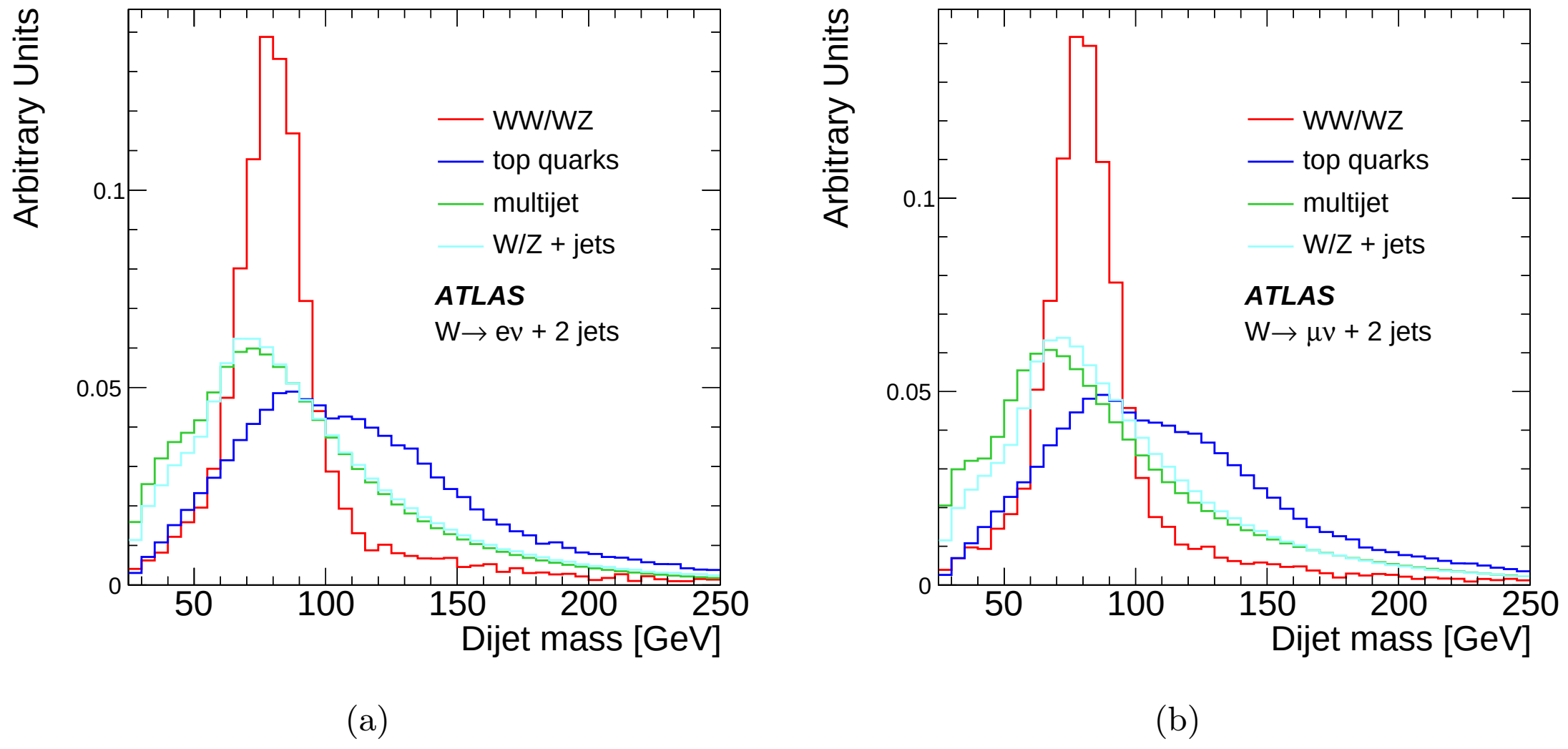

**Figure 2**. The nominal templates for the reconstructed dijet invariant mass for (a) the electron and (b) the muon channels. The templates for $WW/WZ$, $W/Z+$jets and top quarks, including single-top production, are obtained from MC, while the multijet template is obtained using a data-driven method. All templates are normalised to unit area.

where $C_{\ell}^{WV}$ are the ratios of the detector-level signal yield after all analysis cuts to the signal yield in the fiducial phase space for the respective processes and lepton flavour. The values of $C^{WW}$ and $C^{WZ}$ vary between 0.61 and 0.74 and depend on the process and on the channel (electron, muon) considered. The factor $f_{\text{fid}}^{WW}$ represents the ratio of the $WW$ to the $WW+WZ$ fiducial cross sections. The two processes are not separated by this analysis, so $f_{\text{fid}}^{WW}$ is fixed to the SM value of 0.82, calculated with MC@NLO.

The total cross section is obtained by extrapolating the fiducial event yield to the full phase space using eq. (2.2). The factors $D_{\text{tot},\ell}$ are obtained from theoretical predictions and are defined as:

$$D_{\text{tot},\ell} = f_{\text{tot}}^{WW} \cdot C_{\ell}^{WW} \cdot \mathcal{B}_{\ell}^{WW} \cdot A_{\ell}^{WW} + (1 - f_{\text{tot}}^{WW}) \cdot C_{\ell}^{WZ} \cdot \mathcal{B}_{\ell}^{WZ} \cdot A_{\ell}^{WZ} \tag{8.3}$$

where the acceptances $A_{\ell}^{WW}$ and $A_{\ell}^{WZ}$ are calculated as the fraction of signal events satisfying the fiducial-volume selection criteria; they vary in the range 0.08–0.09 depending on the process and are independent of the lepton flavour. $\mathcal{B}_{\ell}^{WW}$ and $\mathcal{B}_{\ell}^{WZ}$ are the branching fractions for the decays $WW \to \ell\nu jj$ and $WZ \to \ell\nu jj$ respectively [54].

## 9 Systematic uncertainties

The total systematic uncertainties on the fiducial and total cross sections are obtained by summing in quadrature the uncertainties on the signal yield, on the factors $D_{\text{fid}}$ or $D_{\text{tot}}$, and on the integrated luminosity.

Systematic uncertainties that affect the fitted signal yield are accounted for by including nuisance parameters with Gaussian constraints in the maximum-likelihood fit ("profiled" systematic uncertainties), with a few exceptions that are described below. The nuisance parameters describe the estimated rate or shape variations of the templates for the various processes. Systematic uncertainties arising from the same source are assumed to be 100% correlated between the electron and muon channels. Uncertainties from different sources are assumed to be independent.

Two of the largest systematic uncertainties are the jet energy scale (JES) and jet energy resolution (JER) uncertainties, determined as described in refs. [51] and [55]. The JES uncertainty also includes the effect of energy deposits due to pile-up, and the uncertainties on the JES and JER are propagated to the $E_{\mathrm{T}}^{\mathrm{miss}}$. The main impact of the JES and JER uncertainties on the measurement of the signal yield is due to the effect of these uncertainties on the shapes of the background distributions.

The largest contribution to the background is from the production of a $W$ or $Z$ boson in association with jets; this background was modelled using ALPGEN. Variations of the factorisation and normalisation scales are considered in evaluating the systematic uncertainty; also, the parameters that describe the matching scheme in the matrix element to initial/final-state radiation (ISR/FSR) particles are varied. Alternative $W/Z+$jets samples generated with SHERPA [27] were also analysed; the $m_{\mathrm{jj}}$ and $p_{\mathrm{T}jj}$ distributions from these samples are consistent with the ALPGEN samples within the aforementioned ALPGEN generator uncertainties, so no additional systematic uncertainty is assigned for ALPGEN-SHERPA differences. The total rate uncertainty assigned to the $W/Z+$jets processes is 20% and it includes rate changes due to cross-section, MC modelling, JES, and JER uncertainties.

The uncertainties on the modelling of the $t\bar{t}$ and single-top processes include shape and rate uncertainties due to variation of the ISR/FSR description. These are calculated with dedicated samples generated with ACERMC. The total rate uncertainty assigned to the single-top and $t\bar{t}$ processes is 15% and includes contributions from cross-section, MC modelling, JES, and JER uncertainties.

The multijet rate and shape uncertainties are determined by cross-checking the data-driven multijet estimation method in a multijet validation region with modified selection criteria: $E_{\mathrm{T}}^{\mathrm{miss}} < 25$ GeV and $m_{\mathrm{T}} > 10$ GeV. The uncertainty on the multijet rate is determined to be 15%. Shape and rate uncertainties for the electron and muon channels are assumed to be uncorrelated. The shape uncertainties are described in the likelihood fit by means of two independent nuisance parameters, one for the electron channel and one for the muon channel. The effect of the multijet rate uncertainty on the extracted signal yield is estimated using pseudo-experiments as mentioned below.

The signal shape modelling uncertainty (including sources such as fragmentation, parton-shower, underlying-event and hadronisation modelling) is assessed by considering alternative templates obtained with samples produced with the PYTHIA generator. Varying the PDF is found to have a negligible impact on the shape of the $m_{\mathrm{jj}}$ and $p_{\mathrm{T}jj}$ distributions.

Some uncertainties on the fitted signal yield were not described through nuisance parameters, either in order to limit the number of parameters in the fit, or because of the difficulty of fully parameterising the possible systematic variation in terms of a nuisance parameter. In such cases the impact of these uncertainties on the signal yield is estimated using an ensemble of pseudo-experiments. These uncertainties include the multijet rate uncertainty and the uncertainty due to the size of the MC event samples. The finite size of the MC event samples produces an uncertainty since it limits the precision with which the $m_{\mathrm{jj}}$ templates are known. This systematic uncertainty is one of the largest, and is dominated by the size of the event sample for the $W+$jets process.

The total uncertainty on the signal yield is obtained by summing contributions from the profiled and non-profiled sources in quadrature.

The fiducial and total cross sections are also affected by uncertainties on the values of $D_{\mathrm{fid}}$ and $D_{\mathrm{tot}}$, respectively. The following sources of uncertainty are considered for these factors: JES, JER, PDF, signal modelling (fragmentation, underlying-event, parton-shower, hadronisation, loss of spin-correlation information), lepton trigger and reconstruction efficiencies, and lepton energy scale. The largest contributions to the $D_{\mathrm{fid}}$ and $D_{\mathrm{tot}}$ uncertainties come from the JES and JER uncertainties while the uncertainties affecting the leptons give very small contributions.

Table 2 summarizes the percent contributions to the systematic uncertainties on the cross sections from the different sources. In the case of profiled systematic uncertainties, the contribution of each individual source to the total uncertainty on $N_{\ell}^{WV}$ is estimated by repeating the fit while fixing the nuisance parameter associated with the source under consideration to its best-fit value. The uncertainty on $N_{\ell}^{WV}$ from this modified fit is subtracted in quadrature from the uncertainty on $N_{\ell}^{WV}$ given by the nominal fit, and the result is taken to be the systematic uncertainty due to the source in question. The data statistics uncertainty is calculated as the fit uncertainty on $N_{\ell}^{WV}$ when all nuisance parameters are fixed to their best-fit values. The largest source of uncertainty is the $W/Z+$jets rate, dominated by the $W+$jets rate uncertainty.

## 10 Cross-section results

The $m_{\mathrm{jj}}$ maximum-likelihood fit, including all the nuisance parameters, is performed on the data, and yields a value of $\beta = 1.11 \pm 0.26$, where $\beta$ is defined in eq. (8.1). The uncertainty includes all the systematic uncertainties from the profiled sources; the purely statistical uncertainty on $\beta$ is 10%. The total systematic uncertainty on the signal yield, including unprofiled systematic uncertainties, is 26%. The measured signal yields are $N_{e}^{WV} = 1970 \pm 200$ (stat.) $\pm$ 500 (syst.) and $N_{\mu}^{WV} = 2190 \pm 220$ (stat.) $\pm$ 560 (syst.) in the electron and muon channels respectively. This signal yield translates into a fiducial cross section of

$$\sigma_{\mathrm{fid}} = 1.37 \pm 0.14\ (\mathrm{stat.}) \pm 0.37\ (\mathrm{syst.})\ \mathrm{pb} \tag{10.1}$$

for the $WW$ and $WZ$ production processes summed over the muon and electron channels, and a total cross section of

$$\sigma_{\mathrm{tot}} = 68 \pm 7\ (\mathrm{stat.}) \pm 19\ (\mathrm{syst.})\ \mathrm{pb}\,, \tag{10.2}$$

in good agreement with the Standard Model prediction obtained with MC@NLO of $\sigma_{\mathrm{tot}} = 61.1 \pm 2.2$ pb.

**Table 2**. Statistical and systematic uncertainties, in %, on the measured fiducial and total cross sections. The uncertainties are split according to the quantity ($N_\ell^{WV}, D_{\rm fid}, D_{\rm tot}, \mathcal{L}$) they are affecting.

| Source | $\sigma_{\rm fid}$ | $\sigma_{\rm tot}$ |
|---|---|---|
| | $N_\ell^{WV}$ | |
| Data statistics | ±10 | |
| MC statistics | ±12 | |
| $W/Z$ + jets rate and shape modelling | ±17 | |
| Multijet shape and rate | ±8 | |
| Top rate and initial/final-state radiation shape modelling | ±6 | |
| Jet energy scale (background and signal shapes) | ±9 | |
| Jet energy resolution (background and signal shapes) | ±11 | |
| $WV$ shape modelling | ±5 | |
| | $D_{\rm fid}$ | $D_{\rm tot}$ |
| JES/JER uncertainty | ±6 | ±6 |
| Signal modelling | ±4 | ±5 |
| Jet veto scale dependence | - | ±5 |
| Others (loss of spin-corr information, lepton uncertainties, PDF) | ±1 | ±4 |
| Luminosity | ±1.8 | |
| Total systematic uncertainty | ±27 | ±28 |

The signal yield significance is estimated using the likelihood ratio, defined as the ratio of the maximum-likelihood with the signal fixed to zero, to the maximum-likelihood including the signal component in the fit [56, 57]. The expected significance is estimated to be $3.2\sigma$ by performing fits with and without the signal component to pseudo-data generated from MC samples with and without the signal component. The observed significance is $3.4\sigma$. The effect of systematic uncertainties is included in the significance calculations.

The $m_{\rm jj}$ distribution of the data overlaid with the fit result is shown in figure 3 for the sum of the electron and muon channels. In addition, the background-subtracted data is shown overlaid with the fitted signal distribution.

As a cross-check, separate fits to the electron and muon channels were performed to extract the most probable $\beta$ values for the two channels. The values obtained, $1.00 \pm 0.37$ for the electron channel and $1.13 \pm 0.36$ for the muon channel, are in agreement with the value obtained with the simultaneous fit.

## 11 Anomalous triple gauge couplings

The measured $WV$ cross section agrees well with the SM predictions; in this section limits are set on anomalous triple gauge couplings affecting the $WWZ$ and $WW\gamma$ vertices. Anomalous couplings tend to enhance the diboson cross section at high boson $p_{\rm T}$. Limits on the anomalous couplings are set by fitting the distribution of the transverse momentum of the reconstructed hadronically decaying $V$, $p_{\mathrm{T}jj}$. The event selection is the same as used for the cross-section measurement, except that $m_{\mathrm{jj}}$ is additionally required to be between 75 and 95 GeV to improve the signal-to-background ratio. The $m_{\mathrm{jj}}$ range and the binning of the $p_{\mathrm{T}jj}$ histogram are chosen to optimise the expected aTGC limits.

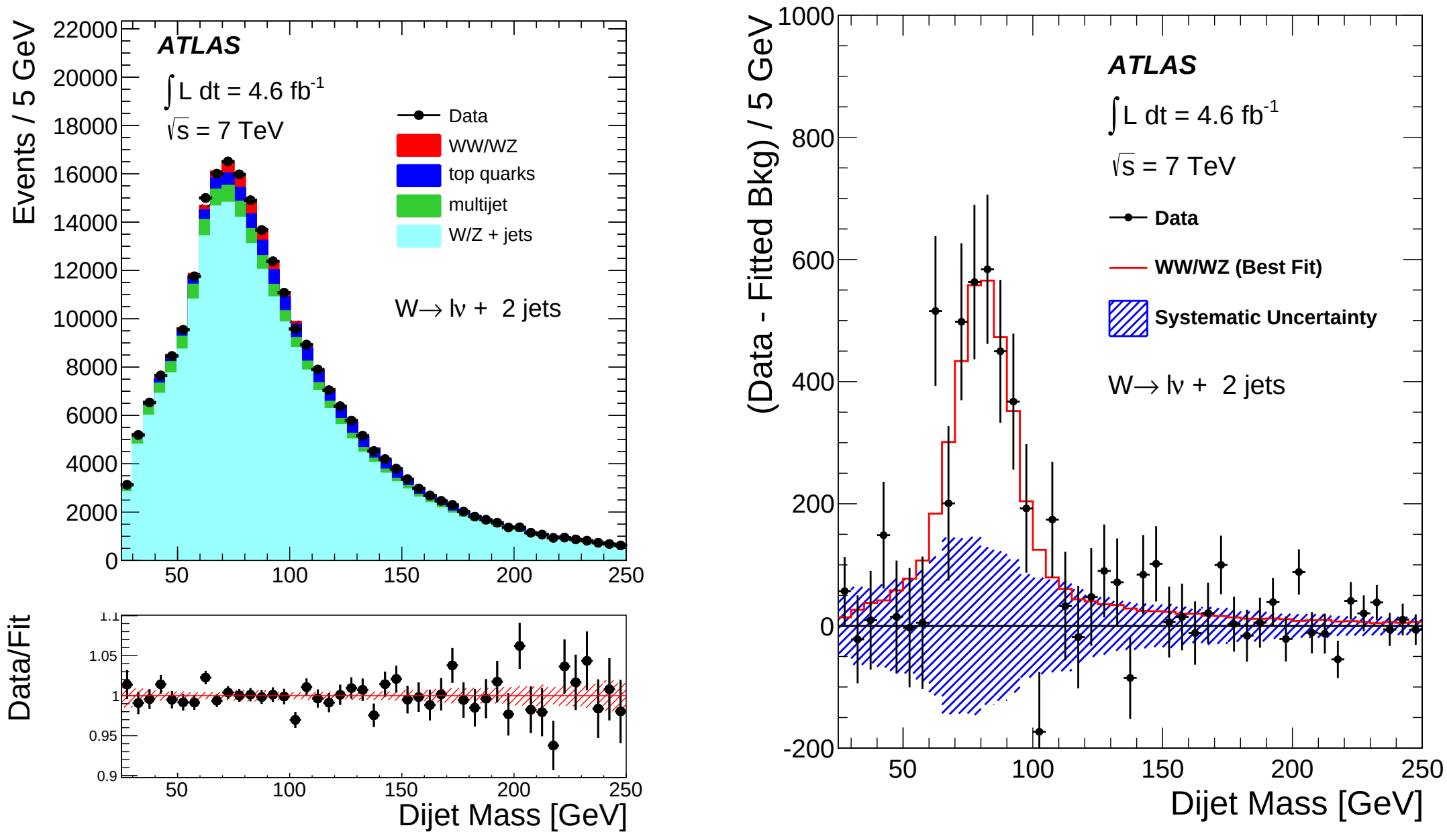

(a) (b)

**Figure 3**. (a) Distributions of the dijet invariant mass for the sum of the electron and muon channels after the likelihood fit. The error bars represent statistical uncertainties, and the stacked histograms are the signal and background contributions. The normalisations and shapes of the histograms are obtained from the best fit to the data, after being allowed to vary within their systematic uncertainties. The lower panel displays the ratio between the data and the total fit result, including both signal and backgrounds. The hatched band shows the systematic uncertainty on the fitted signal plus background. (b) Distribution of the background-subtracted data for the sum of the electron and muon channels. The error bars represent the statistical error on the data. The superimposed histogram shows the fitted signal and the hatched band shows the systematic uncertainty on the background after profiling the nuisance parameters.

To quantify possible deviations from the SM affecting triple gauge boson vertices, the couplings of the $WWZ$ and $WW\gamma$ vertices are described in terms of five dimensionless parameters: $\lambda_\gamma$, $\lambda_Z$, $\kappa_\gamma$, $\kappa_Z$, and $g_1^Z$, only considering couplings that conserve $C$ and $P$ and satisfy electromagnetic gauge invariance [58]. No form factors are applied to these parameters in this analysis. In the SM, $\lambda_\gamma = \lambda_Z = 0$, and $\kappa_\gamma = \kappa_Z = g_1^Z = 1$. Various assumptions can be made to decrease the number of free parameters. In this analysis, limits are given using the so-called LEP scenario [59] in which the following additional constraints, derived from $SU(2) \times U(1)$ gauge invariance, are imposed:

$$\lambda_\gamma = \lambda_Z \equiv \lambda\,, \tag{11.1}$$

$$\Delta\kappa_Z = \Delta g_1^Z - \Delta\kappa_\gamma \tan^2\theta_\mathrm{w}\,, \tag{11.2}$$

where $\Delta\kappa_\gamma \equiv \kappa_\gamma - 1$, $\Delta\kappa_Z \equiv \kappa_Z - 1$, and $\Delta g_1^Z \equiv g_1^Z - 1$. In this scenario, there are three free parameters: $\lambda$, $\Delta\kappa_\gamma$, and $\Delta g_1^Z$.

An alternative approach to the aTGC parametrisation describes deviations from the SM in terms of an effective-field-theory (EFT), valid only up to some mass scale $\Lambda$. This EFT [1, 60] contains three $C$- and $P$-conserving dimension-6 operators. The coefficients of these operators are denoted by $c_W$, $c_B$, and $c_{WWW}$, and can be related to the LEP-scenario parameters by the following equations:

$$\frac{c_W}{\Lambda^2} = \frac{2}{m_Z^2}\Delta g_1^Z\,, \tag{11.3}$$

$$\frac{c_B}{\Lambda^2} = \frac{2}{m_W^2}\Delta\kappa_\gamma - \frac{2}{m_Z^2}\Delta g_1^Z\,, \tag{11.4}$$

$$\frac{c_{WWW}}{\Lambda^2} = \frac{2}{3g^2 m_W^2}\lambda\,, \tag{11.5}$$

where $g$ is the electroweak coupling constant.

The diboson signal with anomalous couplings is modeled using the same generator (MC@NLO+HERWIG) as for the SM signal. The dijet $p_\mathrm{T}$ distribution is shown in figure 4 for data and MC simulation, along with the signal prediction for an aTGC of $\lambda = 0.05$. The limits on the anomalous couplings are calculated by performing a binned maximum-likelihood fit to the $p_{\mathrm{T}jj}$ spectrum. To determine whether a point $\vec{\alpha}$ in the anomalous coupling parameter space is excluded by the data, the likelihood ratio $L(\vec{\alpha})/L(\vec{\alpha}_\mathrm{max})$ is computed, where $\vec{\alpha}_\mathrm{max}$ is the value of the anomalous coupling(s) that maximizes the likelihood. Then the probability of observing such a small likelihood ratio is determined through pseudo-experiments, in which pseudo-data are generated by randomly sampling the probability density function. Systematic uncertainties are incorporated in the fit via nuisance parameters which affect the rates and $p_{\mathrm{T}jj}$ distribution shapes of the signal and background processes. The same sources of systematic uncertainty are included as are described for the $m_\mathrm{jj}$ fit in section 9, except for those found to be negligible, such as the effect of PDF uncertainties on the signal. In addition, an uncertainty is included on the $p_{\mathrm{T}jj}$ distribution shape of the signal due to increasing and decreasing the scales by a factor of two. The factorisation and renormalisation scales are varied simultaneously by the same amount. As can be seen in figure 4, at very high $p_{\mathrm{T}jj}$ the statistical uncertainties dominate, whereas at lower values of $p_{\mathrm{T}jj}$ the systematic uncertainties are more important.

The expected and observed 95% CL limits for $\lambda$, $\Delta\kappa_\gamma$, and $\Delta g_1^Z$ in the LEP scenario are given in table 3. If there were no systematic uncertainties at all, the expected aTGC limits would improve by about 25%.

In figure 5, the observed limits are compared with previous limits from ATLAS [3, 4, 61], CMS [6, 9, 62], D0 [11], and LEP [63], in a variety of channels including $WW \to \ell\nu\ell\nu$, $WZ \to \ell\nu\ell\ell$, $WV \to \ell\nu\mathrm{jj}$, and $W\gamma \to \ell\nu\gamma$. All limits are given at 95% CL, and calculated within the LEP scenario. The form factor $\Lambda_{\text{FF}}$ used for each limit calculation is specified on the figure 5; $\Lambda_{\text{FF}} = \infty$ is equivalent to no form factor. The limits for each parameter are obtained while fixing the other two parameters to zero. In the CMS $\ell\nu$jj analysis and the in ATLAS and CMS $W\gamma$ analyses, no limits on $\Delta g_1^Z$ were given. The ATLAS $WW$ and $WZ$ analyses gave limits on $\Delta g_1^Z$, but with $\Delta\kappa_Z = 0$ rather than $\Delta\kappa_\gamma = 0$, so they are not comparable with these results and are thus excluded. For the ATLAS $WW$ result, the published limits on $\Delta\kappa_Z$ are converted to limits on $\Delta\kappa_\gamma$ using the formula $\Delta\kappa_Z = -\Delta\kappa_\gamma \tan^2\theta_{\text{w}}$. The ATLAS $WZ$ analysis published limits on $\Delta\kappa_Z$, which can also be converted to $\Delta\kappa_\gamma$, but those limits are not shown, since they are much larger than the other limits in this figure. The limits obtained in this analysis are competitive with the limits from the other analyses. Compared to the fully leptonic $WW$ analyses from hadron colliders, the limits shown here are slightly more stringent for $\lambda$ and $\Delta g_1^Z$ and slightly worse for $\Delta\kappa_\gamma$.

**Table 3**. The observed and expected 95% CL limits on the anomalous triple gauge coupling parameters $\lambda$, $\Delta\kappa_\gamma$, and $\Delta g_1^Z$ in the LEP scenario with no form factor applied. The limits on each parameter are calculated while fixing the other two parameters to zero.

| Parameter | Observed Limit | Expected Limit |
|---|---|---|
| $\lambda_Z = \lambda_\gamma$ | $[-0.039, 0.040]$ | $[-0.048, 0.047]$ |
| $\Delta\kappa_\gamma$ | $[-0.21, 0.22]$ | $[-0.23, 0.25]$ |
| $\Delta g_1^Z$ | $[-0.055, 0.071]$ | $[-0.072, 0.085]$ |

In table 4, the limits are shown for each of the five aTGC parameters when no relationship between the different parameters is imposed. In this scenario, $\Delta g_1^Z$ has very little effect on the $WW$ process, whereas $\Delta\kappa_Z$ has very little effect on the $WZ$ process.

**Table 4**. The observed and expected 95% CL limits on the anomalous triple gauge parameters $\lambda_Z$, $\Delta\kappa_Z$, $\Delta g_1^Z$, $\lambda_\gamma$, and $\Delta\kappa_\gamma$, not subjected to any constraints between them. No form factors are applied to the aTGC parameters. The limits on each parameter are calculated while fixing the other four parameters to zero.

| Parameter | Observed Limit | Expected Limit |
|---|---|---|
| $\lambda_Z$ | $[-0.043, 0.044]$ | $[-0.056, 0.056]$ |
| $\Delta\kappa_Z$ | $[-0.090, 0.105]$ | $[-0.11, 0.12]$ |
| $\Delta g_1^Z$ | $[-0.073, 0.095]$ | $[-0.11, 0.12]$ |
| $\lambda_\gamma$ | $[-0.15, 0.15]$ | $[-0.17, 0.16]$ |
| $\Delta\kappa_\gamma$ | $[-0.19, 0.23]$ | $[-0.22, 0.25]$ |

Thus, analyses that restrict themselves to either the $WW$ process or the $WZ$ process have limited sensitivity to at least one of the aTGC parameters. In contrast, this analysis combines the two processes, and therefore has good sensitivity to all five aTGC parameters. As an illustration, this analysis has four times better expected limits on $\Delta g_1^Z$ than the ATLAS $WW \to \ell\nu\ell\nu$ analysis [3], and four times better expected limits on $\Delta\kappa_Z$ than the ATLAS $WZ \to \ell\nu\ell\ell$ analysis [4].

Finally, table 5, gives limits on the EFT parameters. The limits on the EFT parameters $c_W$, $c_B$, and $c_{WWW}$ are in the range $(10\text{–}70)\times(\Lambda/\mathrm{TeV})^2$. In all cases, when computing the limits on one parameter, all the other parameters are fixed to zero.

**Table 5**. The observed and expected 95% CL limits on the effective field theory parameters $c_{WWW}/\Lambda^2$, $c_B/\Lambda^2$, and $c_W/\Lambda^2$. The limits on each parameter are calculated while fixing the other two parameters to zero.

| Parameter | Observed Limit | Expected Limit |
|---|---|---|
| $c_{WWW}/\Lambda^2$ | $[-9.5, 9.6]$ TeV$^{-2}$ | $[-11.6, 11.5]$ TeV$^{-2}$ |
| $c_B/\Lambda^2$ | $[-64, 69]$ TeV$^{-2}$ | $[-73, 79]$ TeV$^{-2}$ |
| $c_W/\Lambda^2$ | $[-13, 18]$ TeV$^{-2}$ | $[-17, 21]$ TeV$^{-2}$ |

The observed two-dimensional 95% CL limits are shown in figure 6 for the LEP scenario. The limits on $\Delta\kappa_\gamma$ and $\Delta g_1^Z$ are significantly correlated, but the limits on the other pairs of parameters do not have large correlations. In addition, the observed two-dimensional 95% CL limits on the EFT parameters are shown in figure 7. None of the EFT parameter pairs exhibit strong correlations.

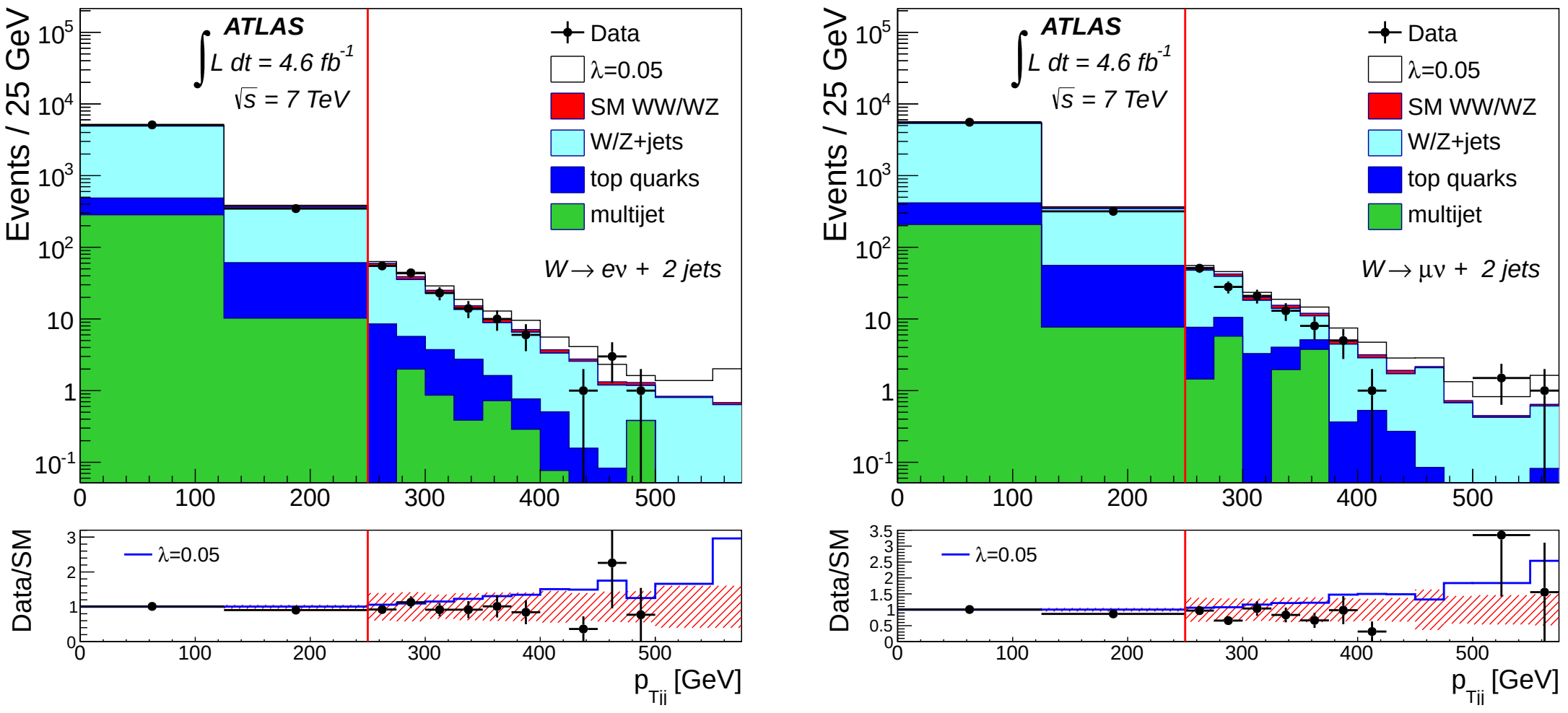

(a) (b)

**Figure 4**. The observed distribution of the transverse momentum of the two jets, compared to the expectation for SM signal plus background, for (a) the electron channel and (b) the muon channel. The error bars represent statistical uncertainties, and the stacked histograms are background and signal predictions as described in the legend. The effect of an aTGC of $\lambda_Z = \lambda_\gamma = 0.05$ is shown for comparison (white histogram) on top of the SM predictions (coloured histograms). The rightmost bin includes overflow. The bottom panels show the ratio between the data and the SM prediction overlaid with the systematic uncertainty on the shape of the $p_{\mathrm{T}jj}$ distribution. The binning in the plots is the same as that one used to perform the calculation of the limits. The red vertical line indicates that the event selection is different for $p_{\mathrm{T}jj}$ less than and greater than 250 GeV, as described in section 6.

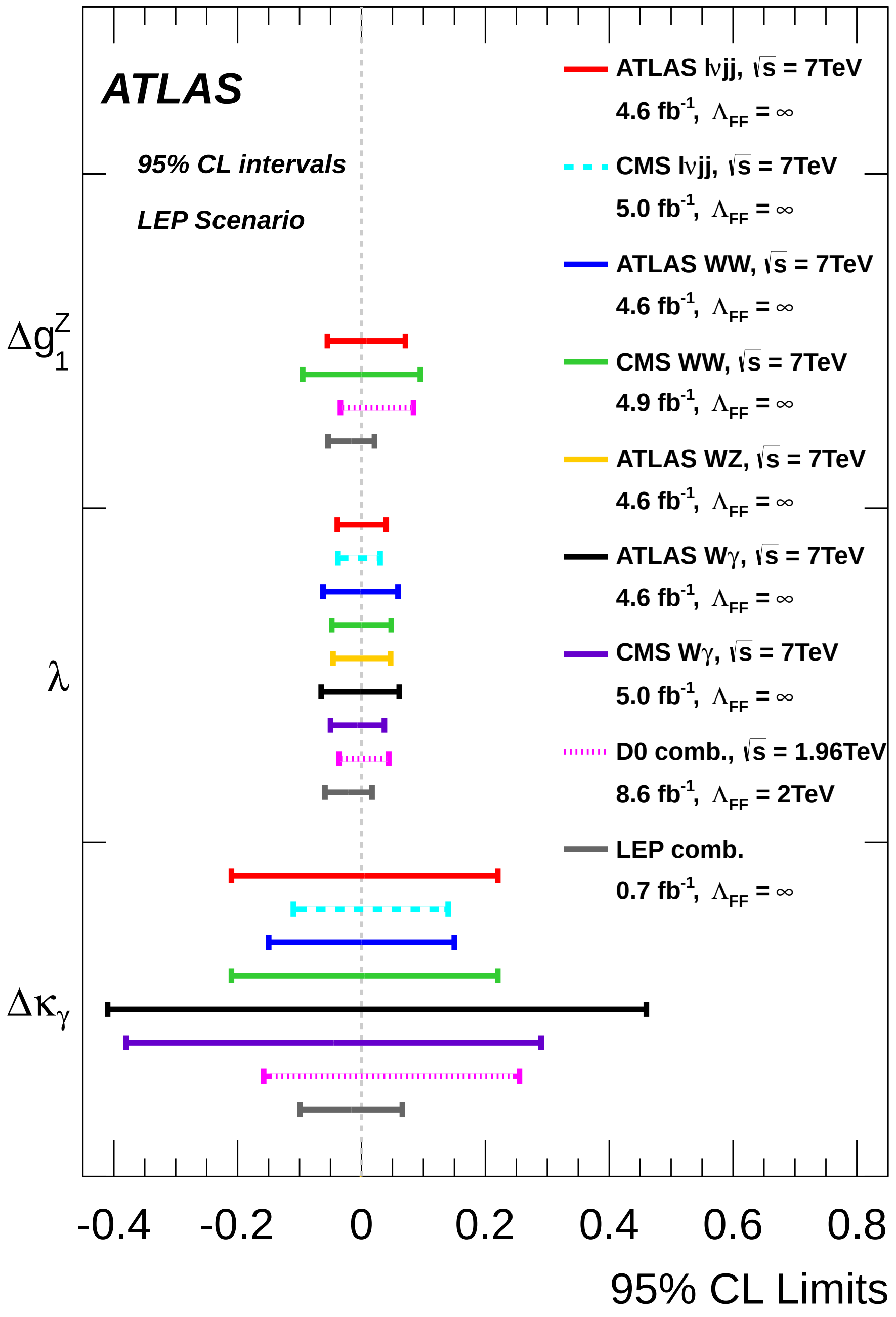

**Figure 5**. Comparison of limits on anomalous triple gauge coupling parameters obtained in this analysis with limits quoted by other experiments and/or in different channels (see text for details).

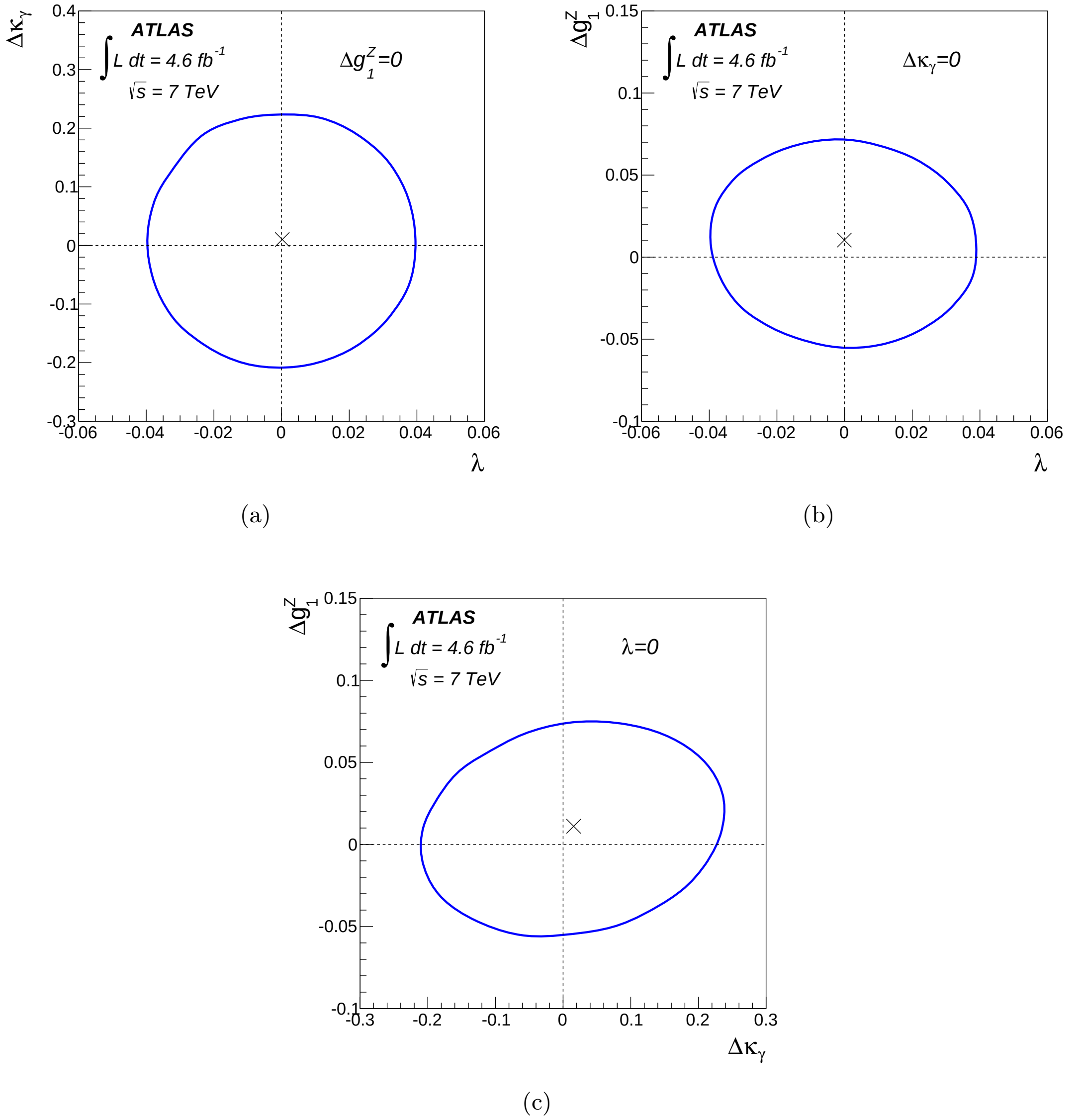

**Figure 6**. The observed two-dimensional 95% CL contours for the anomalous triple gauge couplings (a) $\lambda$ versus $\Delta\kappa_\gamma$, (b) $\lambda$ versus $\Delta g_1^Z$, and (c) $\Delta\kappa_\gamma$ versus $\Delta g_1^Z$. The limits are for the LEP scenario without a form factor.

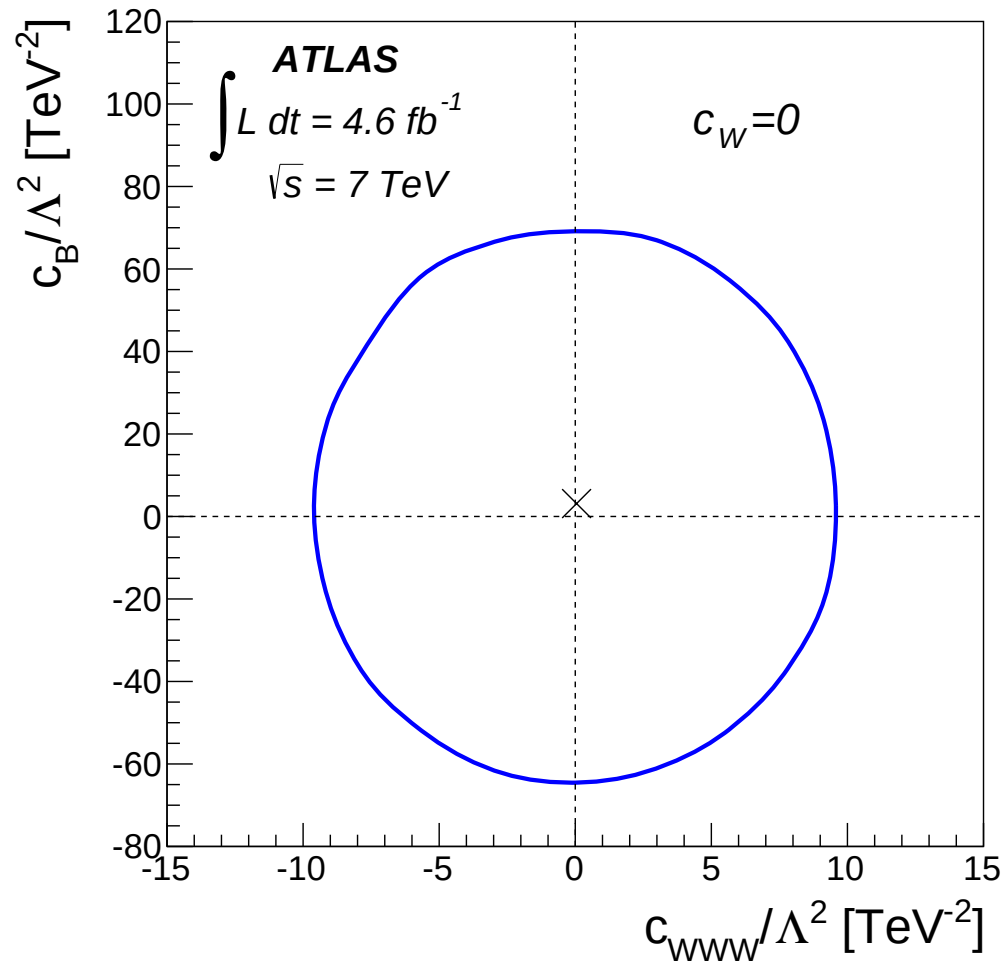

(a)

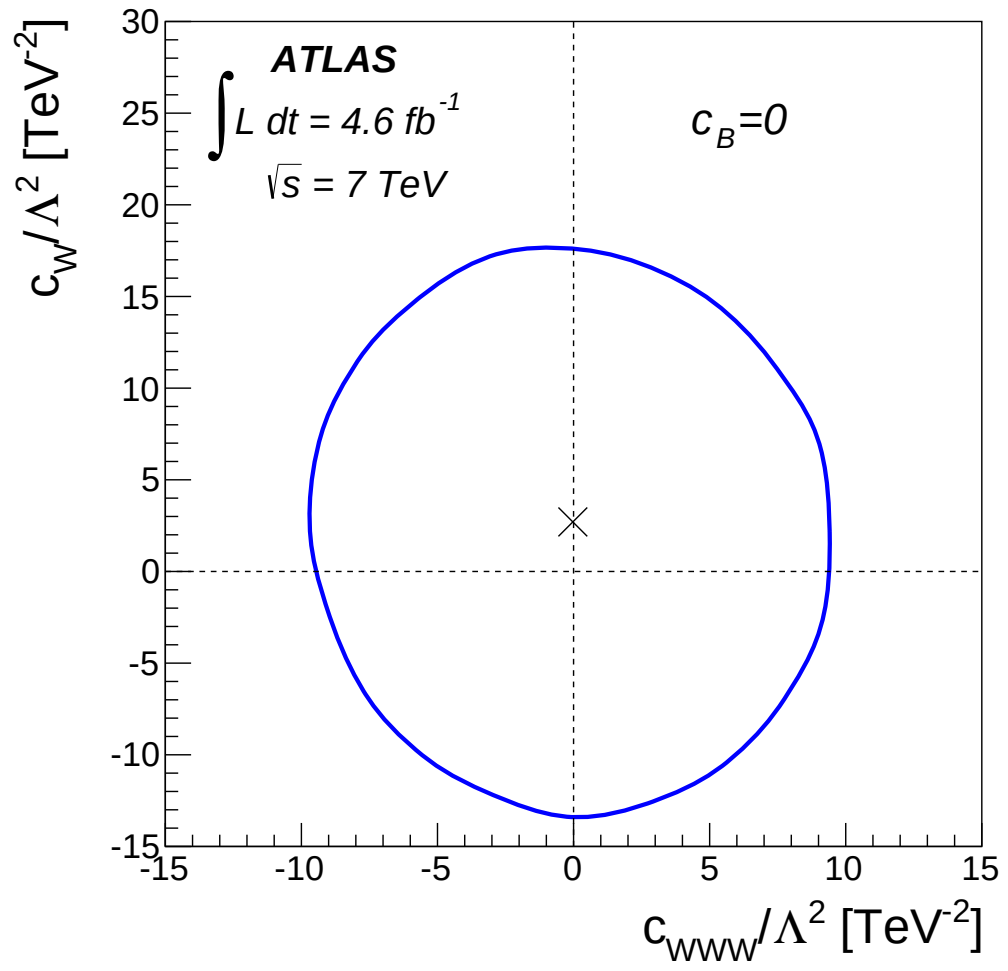

(b)

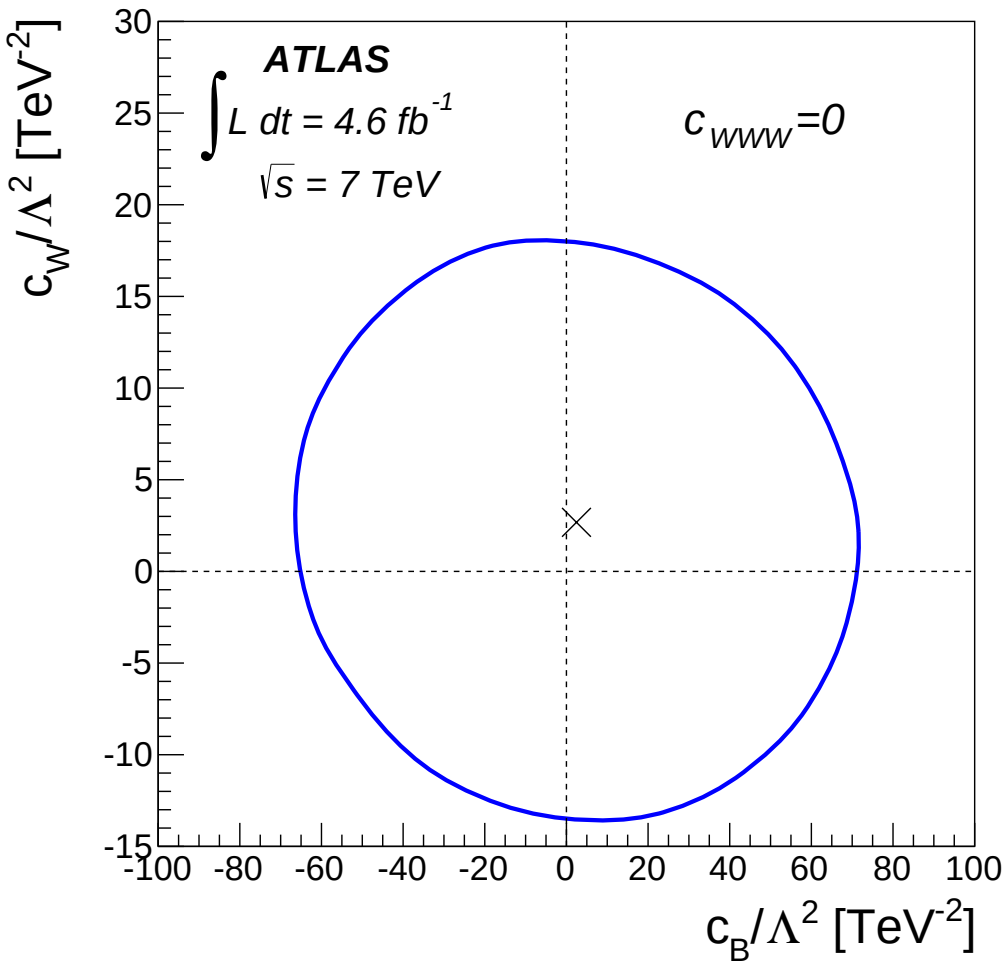

(c)

**Figure 7**. The observed two-dimensional 95% CL contours for the effective field theory parameters (a) $c_{WWW}/\Lambda^2$ versus $c_B/\Lambda^2$, (b) $c_{WWW}/\Lambda^2$ versus $c_W/\Lambda^2$, and (c) $c_B/\Lambda^2$ versus $c_W/\Lambda^2$.

## 12 Conclusion

A measurement of the $pp \to WV$ cross section ($V = W, Z$) at $\sqrt{s} = 7$ TeV is performed with $4.6 \pm 0.1$ fb$^{-1}$ of data collected by ATLAS at the LHC, using the $WV \to \ell\nu$jj ($\ell$=e,$\mu$) decay channels. The total $WW + WZ$ cross section is measured to be $\sigma(WW + WZ) = 68 \pm 7$ (stat.) $\pm$ 19 (syst.) pb, where the observed significance of the signal is $3.4\sigma$. This measurement is consistent with the MC@NLO cross-section prediction of $61.1 \pm 2.2$ pb. In addition, a fiducial cross section is measured in a phase space corresponding closely to the event selection used in the analysis, and is found to be $\sigma_{\text{fid}} = 1.37 \pm 0.14$ (stat.) $\pm$ 0.37 (syst.) pb.

The same process is also used to place limits on anomalous triple gauge couplings (aTGCs) and on the coefficients of dimension-6 operators of an effective-field-theory. Within the LEP scenario, the observed 95% CL limits on the anomalous triple gauge parameters are $-0.039 < \lambda < 0.040$, $-0.21 < \Delta\kappa_\gamma < 0.22$, and $-0.055 < \Delta g_1^Z < 0.071$. The limits on anomalous couplings are similar to those obtained by other diboson analyses.

## Acknowledgments

We thank CERN for the very successful operation of the LHC, as well as the support staff from our institutions without whom ATLAS could not be operated efficiently.

We acknowledge the support of ANPCyT, Argentina; YerPhI, Armenia; ARC, Australia; BMWFW and FWF, Austria; ANAS, Azerbaijan; SSTC, Belarus; CNPq and FAPESP, Brazil; NSERC, NRC and CFI, Canada; CERN; CONICYT, Chile; CAS, MOST and NSFC, China; COLCIENCIAS, Colombia; MSMT CR, MPO CR and VSC CR, Czech Republic; DNRF, DNSRC and Lundbeck Foundation, Denmark; EPLANET, ERC and NSRF, European Union; IN2P3-CNRS, CEA-DSM/IRFU, France; GNSF, Georgia; BMBF, DFG, HGF, MPG and AvH Foundation, Germany; GSRT and NSRF, Greece; ISF, MINERVA, GIF, I-CORE and Benoziyo Center, Israel; INFN, Italy; MEXT and JSPS, Japan; CNRST, Morocco; FOM and NWO, Netherlands; BRF and RCN, Norway; MNiSW and NCN, Poland; GRICES and FCT, Portugal; MNE/IFA, Romania; MES of Russia and ROSATOM, Russian Federation; JINR; MSTD, Serbia; MSSR, Slovakia; ARRS and MIZŠ, Slovenia; DST/NRF, South Africa; MINECO, Spain; SRC and Wallenberg Foundation, Sweden; SER, SNSF and Cantons of Bern and Geneva, Switzerland; NSC, Taiwan; TAEK, Turkey; STFC, the Royal Society and Leverhulme Trust, United Kingdom; DOE and NSF, United States of America.

The crucial computing support from all WLCG partners is acknowledged gratefully, in particular from CERN and the ATLAS Tier-1 facilities at TRIUMF (Canada), NDGF (Denmark, Norway, Sweden), CC-IN2P3 (France), KIT/GridKA (Germany), INFN-CNAF (Italy), NL-T1 (Netherlands), PIC (Spain), ASGC (Taiwan), RAL (UK) and BNL (USA) and in the Tier-2 facilities worldwide.

## The ATLAS Collaboration

G. Aad[85], B. Abbott[113], J. Abdallah[152], S. Abdel Khalek[117], O. Abdinov[11], R. Aben[107], B. Abi[114], M. Abolins[90], O.S. AbouZeid[159], H. Abramowicz[154], H. Abreu[153], R. Abreu[30], Y. Abulaiti[147a,147b], B.S. Acharya[165a,165b,a], L. Adamczyk[38a], D.L. Adams[25], J. Adelman[108], S. Adomeit[100], T. Adye[131], T. Agatonovic-Jovin[13a], J.A. Aguilar-Saavedra[126a,126f], M. Agustoni[17], S.P. Ahlen[22], F. Ahmadov[65,b], G. Aielli[134a,134b], H. Akerstedt[147a,147b], T.P.A. Åkesson[81], G. Akimoto[156], A.V. Akimov[96], G.L. Alberghi[20a,20b], J. Albert[170], S. Albrand[55], M.J. Alconada Verzini[71], M. Aleksa[30], I.N. Aleksandrov[65], C. Alexa[26a], G. Alexander[154], G. Alexandre[49], T. Alexopoulos[10], M. Alhroob[113], G. Alimonti[91a], L. Alio[85], J. Alison[31], B.M.M. Allbrooke[18], L.J. Allison[72], P.P. Allport[74], A. Aloisio[104a,104b], A. Alonso[36], F. Alonso[71], C. Alpigiani[76], A. Altheimer[35], B. Alvarez Gonzalez[90], M.G. Alviggi[104a,104b], K. Amako[66], Y. Amaral Coutinho[24a], C. Amelung[23], D. Amidei[89], S.P. Amor Dos Santos[126a,126c], A. Amorim[126a,126b], S. Amoroso[48], N. Amram[154], G. Amundsen[23], C. Anastopoulos[140], L.S. Ancu[49], N. Andari[30], T. Andeen[35], C.F. Anders[58b], G. Anders[30], K.J. Anderson[31], A. Andreazza[91a,91b], V. Andrei[58a], X.S. Anduaga[71], S. Angelidakis[9], I. Angelozzi[107], P. Anger[44], A. Angerami[35], F. Anghinolfi[30], A.V. Anisenkov[109,c], N. Anjos[12], A. Annovi[47], A. Antonaki[9], M. Antonelli[47], A. Antonov[98], J. Antos[145b], F. Anulli[133a], M. Aoki[66], L. Aperio Bella[18], R. Apolle[120,d], G. Arabidze[90], I. Aracena[144], Y. Arai[66], J.P. Araque[126a], A.T.H. Arce[45], F.A. Arduh[71], J-F. Arguin[95], S. Argyropoulos[42], M. Arik[19a], A.J. Armbruster[30], O. Arnaez[30], V. Arnal[82], H. Arnold[48], M. Arratia[28], O. Arslan[21], A. Artamonov[97], G. Artoni[23], S. Asai[156], N. Asbah[42], A. Ashkenazi[154], B. Åsman[147a,147b], L. Asquith[150], K. Assamagan[25], R. Astalos[145a], M. Atkinson[166], N.B. Atlay[142], B. Auerbach[6], K. Augsten[128], M. Aurousseau[146b], G. Avolio[30], B. Axen[15], G. Azuelos[95,e], Y. Azuma[156], M.A. Baak[30], A.E. Baas[58a], C. Bacci[135a,135b], H. Bachacou[137], K. Bachas[155], M. Backes[30], M. Backhaus[30], E. Badescu[26a], P. Bagiacchi[133a,133b], P. Bagnaia[133a,133b], Y. Bai[33a], T. Bain[35], J.T. Baines[131], O.K. Baker[177], P. Balek[129], F. Balli[137], E. Banas[39], Sw. Banerjee[174], A.A.E. Bannoura[176], H.S. Bansil[18], L. Barak[173], S.P. Baranov[96], E.L. Barberio[88], D. Barberis[50a,50b], M. Barbero[85], T. Barillari[101], M. Barisonzi[176], T. Barklow[144], N. Barlow[28], S.L. Barnes[84], B.M. Barnett[131], R.M. Barnett[15], Z. Barnovska[5], A. Baroncelli[135a], G. Barone[49], A.J. Barr[120], F. Barreiro[82], J. Barreiro Guimarães da Costa[57], R. Bartoldus[144], A.E. Barton[72], P. Bartos[145a], V. Bartsch[150], A. Bassalat[117], A. Basye[166], R.L. Bates[53], S.J. Batista[159], J.R. Batley[28], M. Battaglia[138], M. Battistin[30], F. Bauer[137], H.S. Bawa[144,f], J.B. Beacham[110], M.D. Beattie[72], T. Beau[80], P.H. Beauchemin[162], R. Beccherle[124a,124b], P. Bechtle[21], H.P. Beck[17,g], K. Becker[120], S. Becker[100], M. Beckingham[171], C. Becot[117], A.J. Beddall[19c], A. Beddall[19c], S. Bedikian[177], V.A. Bednyakov[65], C.P. Bee[149], L.J. Beemster[107], T.A. Beermann[176], M. Begel[25], K. Behr[120], C. Belanger-Champagne[87], P.J. Bell[49], W.H. Bell[49], G. Bella[154], L. Bellagamba[20a], A. Bellerive[29], M. Bellomo[86], K. Belotskiy[98], O. Beltramello[30], O. Benary[154], D. Benchekroun[136a], K. Bendtz[147a,147b], N. Benekos[166], Y. Benhammou[154], E. Benhar Noccioli[49], J.A. Benitez Garcia[160b], D.P. Benjamin[45], J.R. Bensinger[23], S. Bentvelsen[107], D. Berge[107],

E. Bergeaas Kuutmann[167], N. Berger[5], F. Berghaus[170], J. Beringer[15], C. Bernard[22],
P. Bernat[78], C. Bernius[110], F.U. Bernlochner[21], T. Berry[77], P. Berta[129], C. Bertella[83],
G. Bertoli[147a,147b], F. Bertolucci[124a,124b], C. Bertsche[113], D. Bertsche[113], M.I. Besana[91a],
G.J. Besjes[106], O. Bessidskaia[147a,147b], M. Bessner[42], N. Besson[137], C. Betancourt[48],
S. Bethke[101], W. Bhimji[46], R.M. Bianchi[125], L. Bianchini[23], M. Bianco[30], O. Biebel[100],
S.P. Bieniek[78], K. Bierwagen[54], M. Biglietti[135a], J. Bilbao De Mendizabal[49], H. Bilokon[47],
M. Bindi[54], S. Binet[117], A. Bingul[19c], C. Bini[133a,133b], C.W. Black[151], J.E. Black[144],
K.M. Black[22], D. Blackburn[139], R.E. Blair[6], J.-B. Blanchard[137], T. Blazek[145a], I. Bloch[42],
C. Blocker[23], W. Blum[83,*], U. Blumenschein[54], G.J. Bobbink[107], V.S. Bobrovnikov[109,c],
S.S. Bocchetta[81], A. Bocci[45], C. Bock[100], C.R. Boddy[120], M. Boehler[48], T.T. Boek[176],
J.A. Bogaerts[30], A.G. Bogdanchikov[109], A. Bogouch[92,*], C. Bohm[147a], V. Boisvert[77],
T. Bold[38a], V. Boldea[26a], A.S. Boldyrev[99], M. Bomben[80], M. Bona[76], M. Boonekamp[137],
A. Borisov[130], G. Borissov[72], M. Borri[84], S. Borroni[42], J. Bortfeldt[100], V. Bortolotto[60a],
K. Bos[107], D. Boscherini[20a], M. Bosman[12], H. Boterenbrood[107], J. Boudreau[125],
J. Bouffard[2], E.V. Bouhova-Thacker[72], D. Boumediene[34], C. Bourdarios[117],
N. Bousson[114], S. Boutouil[136d], A. Boveia[31], J. Boyd[30], I.R. Boyko[65], I. Bozic[13a],
J. Bracinik[18], A. Brandt[8], G. Brandt[15], O. Brandt[58a], U. Bratzler[157], B. Brau[86],
J.E. Brau[116], H.M. Braun[176,*], S.F. Brazzale[165a,165c], B. Brelier[159], K. Brendlinger[122],
A.J. Brennan[88], R. Brenner[167], S. Bressler[173], K. Bristow[146c], T.M. Bristow[46],
D. Britton[53], F.M. Brochu[28], I. Brock[21], R. Brock[90], J. Bronner[101], G. Brooijmans[35],
T. Brooks[77], W.K. Brooks[32b], J. Brosamer[15], E. Brost[116], J. Brown[55],
P.A. Bruckman de Renstrom[39], D. Bruncko[145b], R. Bruneliere[48], S. Brunet[61], A. Bruni[20a],
G. Bruni[20a], M. Bruschi[20a], L. Bryngemark[81], T. Buanes[14], Q. Buat[143], F. Bucci[49],
P. Buchholz[142], A.G. Buckley[53], S.I. Buda[26a], I.A. Budagov[65], F. Buehrer[48], L. Bugge[119],
M.K. Bugge[119], O. Bulekov[98], A.C. Bundock[74], H. Burckhart[30], S. Burdin[74],
B. Burghgrave[108], S. Burke[131], I. Burmeister[43], E. Busato[34], D. Büscher[48], V. Büscher[83],
P. Bussey[53], C.P. Buszello[167], B. Butler[57], J.M. Butler[22], A.I. Butt[3], C.M. Buttar[53],
J.M. Butterworth[78], P. Butti[107], W. Buttinger[28], A. Buzatu[53], M. Byszewski[10],
S. Cabrera Urbán[168], D. Caforio[20a,20b], O. Cakir[4a], P. Calafiura[15], A. Calandri[137],
G. Calderini[80], P. Calfayan[100], L.P. Caloba[24a], D. Calvet[34], S. Calvet[34],
R. Camacho Toro[49], S. Camarda[42], D. Cameron[119], L.M. Caminada[15],
R. Caminal Armadans[12], S. Campana[30], M. Campanelli[78], A. Campoverde[149],
V. Canale[104a,104b], A. Canepa[160a], M. Cano Bret[76], J. Cantero[82], R. Cantrill[126a],
T. Cao[40], M.D.M. Capeans Garrido[30], I. Caprini[26a], M. Caprini[26a], M. Capua[37a,37b],
R. Caputo[83], R. Cardarelli[134a], T. Carli[30], G. Carlino[104a], L. Carminati[91a,91b],
S. Caron[106], E. Carquin[32a], G.D. Carrillo-Montoya[146c], J.R. Carter[28], J. Carvalho[126a,126c],
D. Casadei[78], M.P. Casado[12], M. Casolino[12], E. Castaneda-Miranda[146b], A. Castelli[107],
V. Castillo Gimenez[168], N.F. Castro[126a], P. Catastini[57], A. Catinaccio[30], J.R. Catmore[119],
A. Cattai[30], G. Cattani[134a,134b], J. Caudron[83], V. Cavaliere[166], D. Cavalli[91a],
M. Cavalli-Sforza[12], V. Cavasinni[124a,124b], F. Ceradini[135a,135b], B.C. Cerio[45], K. Cerny[129],
A.S. Cerqueira[24b], A. Cerri[150], L. Cerrito[76], F. Cerutti[15], M. Cerv[30], A. Cervelli[17],
S.A. Cetin[19b], A. Chafaq[136a], D. Chakraborty[108], I. Chalupkova[129], P. Chang[166],
B. Chapleau[87], J.D. Chapman[28], D. Charfeddine[117], D.G. Charlton[18], C.C. Chau[159],

C.A. Chavez Barajas[150], S. Cheatham[153], A. Chegwidden[90], S. Chekanov[6], S.V. Chekulaev[160a], G.A. Chelkov[65,h], M.A. Chelstowska[89], C. Chen[64], H. Chen[25], K. Chen[149], L. Chen[33d,i], S. Chen[33c], X. Chen[33f], Y. Chen[67], H.C. Cheng[89], Y. Cheng[31], A. Cheplakov[65], E. Cheremushkina[130], R. Cherkaoui El Moursli[136e], V. Chernyatin[25,*], E. Cheu[7], L. Chevalier[137], V. Chiarella[47], G. Chiefari[104a,104b], J.T. Childers[6], A. Chilingarov[72], G. Chiodini[73a], A.S. Chisholm[18], R.T. Chislett[78], A. Chitan[26a], M.V. Chizhov[65], S. Chouridou[9], B.K.B. Chow[100], D. Chromek-Burckhart[30], M.L. Chu[152], J. Chudoba[127], J.J. Chwastowski[39], L. Chytka[115], G. Ciapetti[133a,133b], A.K. Ciftci[4a], R. Ciftci[4a], D. Cinca[53], V. Cindro[75], A. Ciocio[15], Z.H. Citron[173], M. Citterio[91a], M. Ciubancan[26a], A. Clark[49], P.J. Clark[46], R.N. Clarke[15], W. Cleland[125], J.C. Clemens[85], C. Clement[147a,147b], Y. Coadou[85], M. Cobal[165a,165c], A. Coccaro[139], J. Cochran[64], L. Coffey[23], J.G. Cogan[144], B. Cole[35], S. Cole[108], A.P. Colijn[107], J. Collot[55], T. Colombo[58c], G. Compostella[101], P. Conde Muiño[126a,126b], E. Coniavitis[48], S.H. Connell[146b], I.A. Connelly[77], S.M. Consonni[91a,91b], V. Consorti[48], S. Constantinescu[26a], C. Conta[121a,121b], G. Conti[57], F. Conventi[104a,j], M. Cooke[15], B.D. Cooper[78], A.M. Cooper-Sarkar[120], N.J. Cooper-Smith[77], K. Copic[15], T. Cornelissen[176], M. Corradi[20a], F. Corriveau[87,k], A. Corso-Radu[164], A. Cortes-Gonzalez[12], G. Cortiana[101], G. Costa[91a], M.J. Costa[168], D. Costanzo[140], D. Côté[8], G. Cottin[28], G. Cowan[77], B.E. Cox[84], K. Cranmer[110], G. Cree[29], S. Crépé-Renaudin[55], F. Crescioli[80], W.A. Cribbs[147a,147b], M. Crispin Ortuzar[120], M. Cristinziani[21], V. Croft[106], G. Crosetti[37a,37b], T. Cuhadar Donszelmann[140], J. Cummings[177], M. Curatolo[47], C. Cuthbert[151], H. Czirr[142], P. Czodrowski[3], S. D'Auria[53], M. D'Onofrio[74], M.J. Da Cunha Sargedas De Sousa[126a,126b], C. Da Via[84], W. Dabrowski[38a], A. Dafinca[120], T. Dai[89], O. Dale[14], F. Dallaire[95], C. Dallapiccola[86], M. Dam[36], A.C. Daniells[18], M. Danninger[169], M. Dano Hoffmann[137], V. Dao[48], G. Darbo[50a], S. Darmora[8], J. Dassoulas[74], A. Dattagupta[61], W. Davey[21], C. David[170], T. Davidek[129], E. Davies[120,d], M. Davies[154], O. Davignon[80], A.R. Davison[78], P. Davison[78], Y. Davygora[58a], E. Dawe[143], I. Dawson[140], R.K. Daya-Ishmukhametova[86], K. De[8], R. de Asmundis[104a], S. De Castro[20a,20b], S. De Cecco[80], N. De Groot[106], P. de Jong[107], H. De la Torre[82], F. De Lorenzi[64], L. De Nooij[107], D. De Pedis[133a], A. De Salvo[133a], U. De Sanctis[150], A. De Santo[150], J.B. De Vivie De Regie[117], W.J. Dearnaley[72], R. Debbe[25], C. Debenedetti[138], B. Dechenaux[55], D.V. Dedovich[65], I. Deigaard[107], J. Del Peso[82], T. Del Prete[124a,124b], F. Deliot[137], C.M. Delitzsch[49], M. Deliyergiyev[75], A. Dell'Acqua[30], L. Dell'Asta[22], M. Dell'Orso[124a,124b], M. Della Pietra[104a,j], D. della Volpe[49], M. Delmastro[5], P.A. Delsart[55], C. Deluca[107], D.A. DeMarco[159], S. Demers[177], M. Demichev[65], A. Demilly[80], S.P. Denisov[130], D. Derendarz[39], J.E. Derkaoui[136d], F. Derue[80], P. Dervan[74], K. Desch[21], C. Deterre[42], P.O. Deviveiros[30], A. Dewhurst[131], S. Dhaliwal[107], A. Di Ciaccio[134a,134b], L. Di Ciaccio[5], A. Di Domenico[133a,133b], C. Di Donato[104a,104b], A. Di Girolamo[30], B. Di Girolamo[30], A. Di Mattia[153], B. Di Micco[135a,135b], R. Di Nardo[47], A. Di Simone[48], R. Di Sipio[20a,20b], D. Di Valentino[29], F.A. Dias[46], M.A. Diaz[32a], E.B. Diehl[89], J. Dietrich[16], T.A. Dietzsch[58a], S. Diglio[85], A. Dimitrievska[13a], J. Dingfelder[21], P. Dita[26a], S. Dita[26a], F. Dittus[30], F. Djama[85], T. Djobava[51b], J.I. Djuvsland[58a], M.A.B. do Vale[24c], D. Dobos[30],

C. Doglioni[49], T. Doherty[53], T. Dohmae[156], J. Dolejsi[129], Z. Dolezal[129],
B.A. Dolgoshein[98,*], M. Donadelli[24d], S. Donati[124a,124b], P. Dondero[121a,121b], J. Donini[34],
J. Dopke[131], A. Doria[104a], M.T. Dova[71], A.T. Doyle[53], M. Dris[10], J. Dubbert[89],
S. Dube[15], E. Dubreuil[34], E. Duchovni[173], G. Duckeck[100], O.A. Ducu[26a], D. Duda[176],
A. Dudarev[30], F. Dudziak[64], L. Duflot[117], L. Duguid[77], M. Dührssen[30], M. Dunford[58a],
H. Duran Yildiz[4a], M. Düren[52], A. Durglishvili[51b], D. Duschinger[44], M. Dwuznik[38a],
M. Dyndal[38a], J. Ebke[100], W. Edson[2], N.C. Edwards[46], W. Ehrenfeld[21], T. Eifert[30],
G. Eigen[14], K. Einsweiler[15], T. Ekelof[167], M. El Kacimi[136c], M. Ellert[167], S. Elles[5],
F. Ellinghaus[83], N. Ellis[30], J. Elmsheuser[100], M. Elsing[30], D. Emeliyanov[131], Y. Enari[156],
O.C. Endner[83], M. Endo[118], R. Engelmann[149], J. Erdmann[177], A. Ereditato[17],
D. Eriksson[147a], G. Ernis[176], J. Ernst[2], M. Ernst[25], J. Ernwein[137], S. Errede[166],
E. Ertel[83], M. Escalier[117], H. Esch[43], C. Escobar[125], B. Esposito[47], A.I. Etienvre[137],
E. Etzion[154], H. Evans[61], A. Ezhilov[123], L. Fabbri[20a,20b], G. Facini[31],
R.M. Fakhrutdinov[130], S. Falciano[133a], R.J. Falla[78], J. Faltova[129], Y. Fang[33a],
M. Fanti[91a,91b], A. Farbin[8], A. Farilla[135a], T. Farooque[12], S. Farrell[15], S.M. Farrington[171],
P. Farthouat[30], F. Fassi[136e], P. Fassnacht[30], D. Fassouliotis[9], A. Favareto[50a,50b],
L. Fayard[117], P. Federic[145a], O.L. Fedin[123,l], W. Fedorko[169], S. Feigl[30], L. Feligioni[85],
C. Feng[33d], E.J. Feng[6], H. Feng[89], A.B. Fenyuk[130], S. Fernandez Perez[30], S. Ferrag[53],
J. Ferrando[53], A. Ferrari[167], P. Ferrari[107], R. Ferrari[121a], D.E. Ferreira de Lima[53],
A. Ferrer[168], D. Ferrere[49], C. Ferretti[89], A. Ferretto Parodi[50a,50b], M. Fiascaris[31],
F. Fiedler[83], A. Filipčič[75], M. Filipuzzi[42], F. Filthaut[106], M. Fincke-Keeler[170],
K.D. Finelli[151], M.C.N. Fiolhais[126a,126c], L. Fiorini[168], A. Firan[40], A. Fischer[2],
J. Fischer[176], W.C. Fisher[90], E.A. Fitzgerald[23], M. Flechl[48], I. Fleck[142], P. Fleischmann[89],
S. Fleischmann[176], G.T. Fletcher[140], G. Fletcher[76], T. Flick[176], A. Floderus[81],
L.R. Flores Castillo[60a], M.J. Flowerdew[101], A. Formica[137], A. Forti[84], D. Fortin[160a],
D. Fournier[117], H. Fox[72], S. Fracchia[12], P. Francavilla[80], M. Franchini[20a,20b],
S. Franchino[30], D. Francis[30], L. Franconi[119], M. Franklin[57], M. Fraternali[121a,121b],
S.T. French[28], C. Friedrich[42], F. Friedrich[44], D. Froidevaux[30], J.A. Frost[120],
C. Fukunaga[157], E. Fullana Torregrosa[83], B.G. Fulsom[144], J. Fuster[168], C. Gabaldon[55],
O. Gabizon[176], A. Gabrielli[20a,20b], A. Gabrielli[133a,133b], S. Gadatsch[107], S. Gadomski[49],
G. Gagliardi[50a,50b], P. Gagnon[61], C. Galea[106], B. Galhardo[126a,126c], E.J. Gallas[120],
B.J. Gallop[131], P. Gallus[128], G. Galster[36], K.K. Gan[111], J. Gao[33b,i], Y.S. Gao[144,f],
F.M. Garay Walls[46], F. Garberson[177], C. García[168], J.E. García Navarro[168],
M. Garcia-Sciveres[15], R.W. Gardner[31], N. Garelli[144], V. Garonne[30], C. Gatti[47],
G. Gaudio[121a], B. Gaur[142], L. Gauthier[95], P. Gauzzi[133a,133b], I.L. Gavrilenko[96], C. Gay[169],
G. Gaycken[21], E.N. Gazis[10], P. Ge[33d], Z. Gecse[169], C.N.P. Gee[131], D.A.A. Geerts[107],
Ch. Geich-Gimbel[21], K. Gellerstedt[147a,147b], C. Gemme[50a], A. Gemmell[53], M.H. Genest[55],
S. Gentile[133a,133b], M. George[54], S. George[77], D. Gerbaudo[164], A. Gershon[154],
H. Ghazlane[136b], N. Ghodbane[34], B. Giacobbe[20a], S. Giagu[133a,133b], V. Giangiobbe[12],
P. Giannetti[124a,124b], F. Gianotti[30], B. Gibbard[25], S.M. Gibson[77], M. Gilchriese[15],
T.P.S. Gillam[28], D. Gillberg[30], G. Gilles[34], D.M. Gingrich[3,e], N. Giokaris[9],
M.P. Giordani[165a,165c], R. Giordano[104a,104b], F.M. Giorgi[20a], F.M. Giorgi[16],
P.F. Giraud[137], D. Giugni[91a], C. Giuliani[48], M. Giulini[58b], B.K. Gjelsten[119],

S. Gkaitatzis[155], I. Gkialas[155], E.L. Gkougkousis[117], L.K. Gladilin[99], C. Glasman[82], J. Glatzer[30], P.C.F. Glaysher[46], A. Glazov[42], G.L. Glonti[62], M. Goblirsch-Kolb[101], J.R. Goddard[76], J. Godlewski[30], C. Goeringer[83], S. Goldfarb[89], T. Golling[177], D. Golubkov[130], A. Gomes[126a,126b,126d], L.S. Gomez Fajardo[42], R. Gonçalo[126a], J. Goncalves Pinto Firmino Da Costa[137], L. Gonella[21], S. González de la Hoz[168], G. Gonzalez Parra[12], S. Gonzalez-Sevilla[49], L. Goossens[30], P.A. Gorbounov[97], H.A. Gordon[25], I. Gorelov[105], B. Gorini[30], E. Gorini[73a,73b], A. Gorišek[75], E. Gornicki[39], A.T. Goshaw[45], C. Gössling[43], M.I. Gostkin[65], M. Gouighri[136a], D. Goujdami[136c], M.P. Goulette[49], A.G. Goussiou[139], C. Goy[5], H.M.X. Grabas[138], L. Graber[54], I. Grabowska-Bold[38a], P. Grafström[20a,20b], K-J. Grahn[42], J. Gramling[49], E. Gramstad[119], S. Grancagnolo[16], V. Grassi[149], V. Gratchev[123], H.M. Gray[30], E. Graziani[135a], O.G. Grebenyuk[123], Z.D. Greenwood[79,m], K. Gregersen[78], I.M. Gregor[42], P. Grenier[144], J. Griffiths[8], A.A. Grillo[138], K. Grimm[72], S. Grinstein[12,n], Ph. Gris[34], Y.V. Grishkevich[99], J.-F. Grivaz[117], J.P. Grohs[44], A. Grohsjean[42], E. Gross[173], J. Grosse-Knetter[54], G.C. Grossi[134a,134b], Z.J. Grout[150], L. Guan[33b], J. Guenther[128], F. Guescini[49], D. Guest[177], O. Gueta[154], C. Guicheney[34], E. Guido[50a,50b], T. Guillemin[117], S. Guindon[2], U. Gul[53], C. Gumpert[44], J. Guo[35], S. Gupta[120], P. Gutierrez[113], N.G. Gutierrez Ortiz[53], C. Gutschow[78], N. Guttman[154], C. Guyot[137], C. Gwenlan[120], C.B. Gwilliam[74], A. Haas[110], C. Haber[15], H.K. Hadavand[8], N. Haddad[136e], P. Haefner[21], S. Hageböck[21], Z. Hajduk[39], H. Hakobyan[178], M. Haleem[42], D. Hall[120], G. Halladjian[90], G.D. Hallewell[85], K. Hamacher[176], P. Hamal[115], K. Hamano[170], M. Hamer[54], A. Hamilton[146a], S. Hamilton[162], G.N. Hamity[146c], P.G. Hamnett[42], L. Han[33b], K. Hanagaki[118], K. Hanawa[156], M. Hance[15], P. Hanke[58a], R. Hanna[137], J.B. Hansen[36], J.D. Hansen[36], P.H. Hansen[36], K. Hara[161], A.S. Hard[174], T. Harenberg[176], F. Hariri[117], S. Harkusha[92], D. Harper[89], R.D. Harrington[46], O.M. Harris[139], P.F. Harrison[171], F. Hartjes[107], M. Hasegawa[67], S. Hasegawa[103], Y. Hasegawa[141], A. Hasib[113], S. Hassani[137], S. Haug[17], M. Hauschild[30], R. Hauser[90], M. Havranek[127], C.M. Hawkes[18], R.J. Hawkings[30], A.D. Hawkins[81], T. Hayashi[161], D. Hayden[90], C.P. Hays[120], J.M. Hays[76], H.S. Hayward[74], S.J. Haywood[131], S.J. Head[18], T. Heck[83], V. Hedberg[81], L. Heelan[8], S. Heim[122], T. Heim[176], B. Heinemann[15], L. Heinrich[110], J. Hejbal[127], L. Helary[22], C. Heller[100], M. Heller[30], S. Hellman[147a,147b], D. Hellmich[21], C. Helsens[30], J. Henderson[120], R.C.W. Henderson[72], Y. Heng[174], C. Hengler[42], A. Henrichs[177], A.M. Henriques Correia[30], S. Henrot-Versille[117], G.H. Herbert[16], Y. Hernández Jiménez[168], R. Herrberg-Schubert[16], G. Herten[48], R. Hertenberger[100], L. Hervas[30], G.G. Hesketh[78], N.P. Hessey[107], R. Hickling[76], E. Higón-Rodriguez[168], E. Hill[170], J.C. Hill[28], K.H. Hiller[42], S.J. Hillier[18], I. Hinchliffe[15], E. Hines[122], M. Hirose[158], D. Hirschbuehl[176], J. Hobbs[149], N. Hod[107], M.C. Hodgkinson[140], P. Hodgson[140], A. Hoecker[30], M.R. Hoeferkamp[105], F. Hoenig[100], D. Hoffmann[85], M. Hohlfeld[83], T.R. Holmes[15], T.M. Hong[122], L. Hooft van Huysduynen[110], W.H. Hopkins[116], Y. Horii[103], A.J. Horton[143], J-Y. Hostachy[55], S. Hou[152], A. Hoummada[136a], J. Howard[120], J. Howarth[42], M. Hrabovsky[115], I. Hristova[16], J. Hrivnac[117], T. Hryn'ova[5], A. Hrynevich[93], C. Hsu[146c], P.J. Hsu[152], S.-C. Hsu[139], D. Hu[35], X. Hu[89], Y. Huang[42], Z. Hubacek[30], F. Hubaut[85], F. Huegging[21],

T.B. Huffman[120], E.W. Hughes[35], G. Hughes[72], M. Huhtinen[30], T.A. Hülsing[83], M. Hurwitz[15], N. Huseynov[65,b], J. Huston[90], J. Huth[57], G. Iacobucci[49], G. Iakovidis[10], I. Ibragimov[142], L. Iconomidou-Fayard[117], E. Ideal[177], Z. Idrissi[136e], P. Iengo[104a], O. Igonkina[107], T. Iizawa[172], Y. Ikegami[66], K. Ikematsu[142], M. Ikeno[66], Y. Ilchenko[31,o], D. Iliadis[155], N. Ilic[159], Y. Inamaru[67], T. Ince[101], P. Ioannou[9], M. Iodice[135a], K. Iordanidou[9], V. Ippolito[57], A. Irles Quiles[168], C. Isaksson[167], M. Ishino[68], M. Ishitsuka[158], R. Ishmukhametov[111], C. Issever[120], S. Istin[19a], J.M. Iturbe Ponce[84], R. Iuppa[134a,134b], J. Ivarsson[81], W. Iwanski[39], H. Iwasaki[66], J.M. Izen[41], V. Izzo[104a], B. Jackson[122], M. Jackson[74], P. Jackson[1], M.R. Jaekel[30], V. Jain[2], K. Jakobs[48], S. Jakobsen[30], T. Jakoubek[127], J. Jakubek[128], D.O. Jamin[152], D.K. Jana[79], E. Jansen[78], H. Jansen[30], J. Janssen[21], M. Janus[171], G. Jarlskog[81], N. Javadov[65,b], T. Javůrek[48], L. Jeanty[15], J. Jejelava[51a,p], G.-Y. Jeng[151], D. Jennens[88], P. Jenni[48,q], J. Jentzsch[43], C. Jeske[171], S. Jézéquel[5], H. Ji[174], J. Jia[149], Y. Jiang[33b], M. Jimenez Belenguer[42], S. Jin[33a], A. Jinaru[26a], O. Jinnouchi[158], M.D. Joergensen[36], K.E. Johansson[147a,147b], P. Johansson[140], K.A. Johns[7], K. Jon-And[147a,147b], G. Jones[171], R.W.L. Jones[72], T.J. Jones[74], J. Jongmanns[58a], P.M. Jorge[126a,126b], K.D. Joshi[84], J. Jovicevic[148], X. Ju[174], C.A. Jung[43], P. Jussel[62], A. Juste Rozas[12,n], M. Kaci[168], A. Kaczmarska[39], M. Kado[117], H. Kagan[111], M. Kagan[144], E. Kajomovitz[45], C.W. Kalderon[120], S. Kama[40], A. Kamenshchikov[130], N. Kanaya[156], M. Kaneda[30], S. Kaneti[28], V.A. Kantserov[98], J. Kanzaki[66], B. Kaplan[110], A. Kapliy[31], D. Kar[53], K. Karakostas[10], A. Karamaoun[3], N. Karastathis[10], M.J. Kareem[54], M. Karnevskiy[83], S.N. Karpov[65], Z.M. Karpova[65], K. Karthik[110], V. Kartvelishvili[72], A.N. Karyukhin[130], L. Kashif[174], G. Kasieczka[58b], R.D. Kass[111], A. Kastanas[14], Y. Kataoka[156], A. Katre[49], J. Katzy[42], V. Kaushik[7], K. Kawagoe[70], T. Kawamoto[156], G. Kawamura[54], S. Kazama[156], V.F. Kazanin[109], M.Y. Kazarinov[65], R. Keeler[170], R. Kehoe[40], M. Keil[54], J.S. Keller[42], J.J. Kempster[77], H. Keoshkerian[5], O. Kepka[127], B.P. Kerševan[75], S. Kersten[176], K. Kessoku[156], J. Keung[159], R.A. Keyes[87], F. Khalil-zada[11], H. Khandanyan[147a,147b], A. Khanov[114], A. Kharlamov[109], A. Khodinov[98], A. Khomich[58a], T.J. Khoo[28], G. Khoriauli[21], V. Khovanskiy[97], E. Khramov[65], J. Khubua[51b], H.Y. Kim[8], H. Kim[147a,147b], S.H. Kim[161], N. Kimura[155], O. Kind[16], B.T. King[74], M. King[168], R.S.B. King[120], S.B. King[169], J. Kirk[131], A.E. Kiryunin[101], T. Kishimoto[67], D. Kisielewska[38a], F. Kiss[48], K. Kiuchi[161], E. Kladiva[145b], M. Klein[74], U. Klein[74], K. Kleinknecht[83], P. Klimek[147a,147b], A. Klimentov[25], R. Klingenberg[43], J.A. Klinger[84], T. Klioutchnikova[30], P.F. Klok[106], E.-E. Kluge[58a], P. Kluit[107], S. Kluth[101], E. Kneringer[62], E.B.F.G. Knoops[85], A. Knue[53], D. Kobayashi[158], T. Kobayashi[156], M. Kobel[44], M. Kocian[144], P. Kodys[129], T. Koffas[29], E. Koffeman[107], L.A. Kogan[120], S. Kohlmann[176], Z. Kohout[128], T. Kohriki[66], T. Koi[144], H. Kolanoski[16], I. Koletsou[5], J. Koll[90], A.A. Komar[96,*], Y. Komori[156], T. Kondo[66], N. Kondrashova[42], K. Köneke[48], A.C. König[106], S. König[83], T. Kono[66,r], R. Konoplich[110,s], N. Konstantinidis[78], R. Kopeliansky[153], S. Koperny[38a], L. Köpke[83], A.K. Kopp[48], K. Korcyl[39], K. Kordas[155], A. Korn[78], A.A. Korol[109,c], I. Korolkov[12], E.V. Korolkova[140], V.A. Korotkov[130], O. Kortner[101], S. Kortner[101], V.V. Kostyukhin[21], V.M. Kotov[65], A. Kotwal[45], A. Kourkoumeli-Charalampidi[155], C. Kourkoumelis[9], V. Kouskoura[25], A. Koutsman[160a], R. Kowalewski[170], T.Z. Kowalski[38a], W. Kozanecki[137],

A.S. Kozhin[130], V.A. Kramarenko[99], G. Kramberger[75], D. Krasnopevtsev[98],
M.W. Krasny[80], A. Krasznahorkay[30], J.K. Kraus[21], A. Kravchenko[25], S. Kreiss[110],
M. Kretz[58c], J. Kretzschmar[74], K. Kreutzfeldt[52], P. Krieger[159], K. Kroeninger[54],
H. Kroha[101], J. Kroll[122], J. Kroseberg[21], J. Krstic[13a], U. Kruchonak[65], H. Krüger[21],
T. Kruker[17], N. Krumnack[64], Z.V. Krumshteyn[65], A. Kruse[174], M.C. Kruse[45],
M. Kruskal[22], T. Kubota[88], H. Kucuk[78], S. Kuday[4c], S. Kuehn[48], A. Kugel[58c], A. Kuhl[138],
T. Kuhl[42], V. Kukhtin[65], Y. Kulchitsky[92], S. Kuleshov[32b], M. Kuna[133a,133b], T. Kunigo[68],
A. Kupco[127], H. Kurashige[67], Y.A. Kurochkin[92], R. Kurumida[67], V. Kus[127],
E.S. Kuwertz[148], M. Kuze[158], J. Kvita[115], D. Kyriazopoulos[140], A. La Rosa[49],
L. La Rotonda[37a,37b], C. Lacasta[168], F. Lacava[133a,133b], J. Lacey[29], H. Lacker[16],
D. Lacour[80], V.R. Lacuesta[168], E. Ladygin[65], R. Lafaye[5], B. Laforge[80], T. Lagouri[177],
S. Lai[48], H. Laier[58a], L. Lambourne[78], S. Lammers[61], C.L. Lampen[7], W. Lampl[7],
E. Lançon[137], U. Landgraf[48], M.P.J. Landon[76], V.S. Lang[58a], A.J. Lankford[164],
F. Lanni[25], K. Lantzsch[30], S. Laplace[80], C. Lapoire[21], J.F. Laporte[137], T. Lari[91a],
F. Lasagni Manghi[20a,20b], M. Lassnig[30], P. Laurelli[47], W. Lavrijsen[15], A.T. Law[138],
P. Laycock[74], O. Le Dortz[80], E. Le Guirriec[85], E. Le Menedeu[12], T. LeCompte[6],
F. Ledroit-Guillon[55], C.A. Lee[146b], H. Lee[107], S.C. Lee[152], L. Lee[1], G. Lefebvre[80],
M. Lefebvre[170], F. Legger[100], C. Leggett[15], A. Lehan[74], G. Lehmann Miotto[30], X. Lei[7],
W.A. Leight[29], A. Leisos[155], A.G. Leister[177], M.A.L. Leite[24d], R. Leitner[129],
D. Lellouch[173], B. Lemmer[54], K.J.C. Leney[78], T. Lenz[21], G. Lenzen[176], B. Lenzi[30],
R. Leone[7], S. Leone[124a,124b], C. Leonidopoulos[46], S. Leontsinis[10], C. Leroy[95],
C.G. Lester[28], C.M. Lester[122], M. Levchenko[123], J. Levêque[5], D. Levin[89],
L.J. Levinson[173], M. Levy[18], A. Lewis[120], G.H. Lewis[110], A.M. Leyko[21], M. Leyton[41],
B. Li[33b,t], B. Li[85], H. Li[149], H.L. Li[31], L. Li[45], L. Li[33e], S. Li[45], Y. Li[33c,u], Z. Liang[138],
H. Liao[34], B. Liberti[134a], P. Lichard[30], K. Lie[166], J. Liebal[21], W. Liebig[14], C. Limbach[21],
A. Limosani[151], S.C. Lin[152,v], T.H. Lin[83], F. Linde[107], B.E. Lindquist[149],
J.T. Linnemann[90], E. Lipeles[122], A. Lipniacka[14], M. Lisovyi[42], T.M. Liss[166],
D. Lissauer[25], A. Lister[169], A.M. Litke[138], B. Liu[152], D. Liu[152], J.B. Liu[33b], K. Liu[33b,w],
L. Liu[89], M. Liu[45], M. Liu[33b], Y. Liu[33b], M. Livan[121a,121b], A. Lleres[55],
J. Llorente Merino[82], S.L. Lloyd[76], F. Lo Sterzo[152], E. Lobodzinska[42], P. Loch[7],
W.S. Lockman[138], F.K. Loebinger[84], A.E. Loevschall-Jensen[36], A. Loginov[177], T. Lohse[16],
K. Lohwasser[42], M. Lokajicek[127], V.P. Lombardo[5], B.A. Long[22], J.D. Long[89],
R.E. Long[72], L. Lopes[126a], D. Lopez Mateos[57], B. Lopez Paredes[140], I. Lopez Paz[12],
J. Lorenz[100], N. Lorenzo Martinez[61], M. Losada[163], P. Loscutoff[15], X. Lou[41],
A. Lounis[117], J. Love[6], P.A. Love[72], A.J. Lowe[144,f], F. Lu[33a], N. Lu[89], H.J. Lubatti[139],
C. Luci[133a,133b], A. Lucotte[55], F. Luehring[61], W. Lukas[62], L. Luminari[133a],
O. Lundberg[147a,147b], B. Lund-Jensen[148], M. Lungwitz[83], D. Lynn[25], R. Lysak[127],
E. Lytken[81], H. Ma[25], L.L. Ma[33d], G. Maccarrone[47], A. Macchiolo[101],
J. Machado Miguens[126a,126b], D. Macina[30], D. Madaffari[85], R. Madar[48], H.J. Maddocks[72],
W.F. Mader[44], A. Madsen[167], M. Maeno[8], T. Maeno[25], A. Maevskiy[99], E. Magradze[54],
K. Mahboubi[48], J. Mahlstedt[107], S. Mahmoud[74], C. Maiani[137], C. Maidantchik[24a],
A.A. Maier[101], A. Maio[126a,126b,126d], S. Majewski[116], Y. Makida[66], N. Makovec[117],
P. Mal[137,x], B. Malaescu[80], Pa. Malecki[39], V.P. Maleev[123], F. Malek[55], U. Mallik[63],

D. Malon[6], C. Malone[144], S. Maltezos[10], V.M. Malyshev[109], S. Malyukov[30], J. Mamuzic[13b], B. Mandelli[30], L. Mandelli[91a], I. Mandić[75], R. Mandrysch[63], J. Maneira[126a,126b], A. Manfredini[101], L. Manhaes de Andrade Filho[24b], J.A. Manjarres Ramos[160b], A. Mann[100], P.M. Manning[138], A. Manousakis-Katsikakis[9], B. Mansoulie[137], R. Mantifel[87], L. Mapelli[30], L. March[146c], J.F. Marchand[29], G. Marchiori[80], M. Marcisovsky[127], C.P. Marino[170], M. Marjanovic[13a], F. Marroquim[24a], S.P. Marsden[84], Z. Marshall[15], L.F. Marti[17], S. Marti-Garcia[168], B. Martin[30], B. Martin[90], T.A. Martin[171], V.J. Martin[46], B. Martin dit Latour[14], H. Martinez[137], M. Martinez[12,n], S. Martin-Haugh[131], A.C. Martyniuk[78], M. Marx[139], F. Marzano[133a], A. Marzin[30], L. Masetti[83], T. Mashimo[156], R. Mashinistov[96], J. Masik[84], A.L. Maslennikov[109,c], I. Massa[20a,20b], L. Massa[20a,20b], N. Massol[5], P. Mastrandrea[149], A. Mastroberardino[37a,37b], T. Masubuchi[156], P. Mättig[176], J. Mattmann[83], J. Maurer[26a], S.J. Maxfield[74], D.A. Maximov[109,c], R. Mazini[152], L. Mazzaferro[134a,134b], G. Mc Goldrick[159], S.P. Mc Kee[89], A. McCarn[89], R.L. McCarthy[149], T.G. McCarthy[29], N.A. McCubbin[131], K.W. McFarlane[56,*], J.A. Mcfayden[78], G. Mchedlidze[54], S.J. McMahon[131], R.A. McPherson[170,k], J. Mechnich[107], M. Medinnis[42], S. Meehan[31], S. Mehlhase[100], A. Mehta[74], K. Meier[58a], C. Meineck[100], B. Meirose[41], C. Melachrinos[31], B.R. Mellado Garcia[146c], F. Meloni[17], A. Mengarelli[20a,20b], S. Menke[101], E. Meoni[162], K.M. Mercurio[57], S. Mergelmeyer[21], N. Meric[137], P. Mermod[49], L. Merola[104a,104b], C. Meroni[91a], F.S. Merritt[31], H. Merritt[111], A. Messina[30,y], J. Metcalfe[25], A.S. Mete[164], C. Meyer[83], C. Meyer[122], J-P. Meyer[137], J. Meyer[30], R.P. Middleton[131], S. Migas[74], S. Miglioranzi[165a,165c], L. Mijović[21], G. Mikenberg[173], M. Mikestikova[127], M. Mikuž[75], A. Milic[30], D.W. Miller[31], C. Mills[46], A. Milov[173], D.A. Milstead[147a,147b], A.A. Minaenko[130], Y. Minami[156], I.A. Minashvili[65], A.I. Mincer[110], B. Mindur[38a], M. Mineev[65], Y. Ming[174], L.M. Mir[12], G. Mirabelli[133a], T. Mitani[172], J. Mitrevski[100], V.A. Mitsou[168], A. Miucci[49], P.S. Miyagawa[140], J.U. Mjörnmark[81], T. Moa[147a,147b], K. Mochizuki[85], S. Mohapatra[35], W. Mohr[48], S. Molander[147a,147b], R. Moles-Valls[168], K. Mönig[42], C. Monini[55], J. Monk[36], E. Monnier[85], J. Montejo Berlingen[12], F. Monticelli[71], S. Monzani[133a,133b], R.W. Moore[3], N. Morange[63], D. Moreno[163], M. Moreno Llácer[54], P. Morettini[50a], M. Morgenstern[44], M. Morii[57], V. Morisbak[119], S. Moritz[83], A.K. Morley[148], G. Mornacchi[30], J.D. Morris[76], A. Morton[42], L. Morvaj[103], H.G. Moser[101], M. Mosidze[51b], J. Moss[111], K. Motohashi[158], R. Mount[144], E. Mountricha[25], S.V. Mouraviev[96,*], E.J.W. Moyse[86], S. Muanza[85], R.D. Mudd[18], F. Mueller[58a], J. Mueller[125], K. Mueller[21], T. Mueller[28], T. Mueller[83], D. Muenstermann[49], Y. Munwes[154], J.A. Murillo Quijada[18], W.J. Murray[171,131], H. Musheghyan[54], E. Musto[153], A.G. Myagkov[130,z], M. Myska[128], O. Nackenhorst[54], J. Nadal[54], K. Nagai[120], R. Nagai[158], Y. Nagai[85], K. Nagano[66], A. Nagarkar[111], Y. Nagasaka[59], K. Nagata[161], M. Nagel[101], A.M. Nairz[30], Y. Nakahama[30], K. Nakamura[66], T. Nakamura[156], I. Nakano[112], H. Namasivayam[41], G. Nanava[21], R.F. Naranjo Garcia[42], R. Narayan[58b], T. Nattermann[21], T. Naumann[42], G. Navarro[163], R. Nayyar[7], H.A. Neal[89], P.Yu. Nechaeva[96], T.J. Neep[84], P.D. Nef[144], A. Negri[121a,121b], G. Negri[30], M. Negrini[20a], S. Nektarijevic[49], C. Nellist[117], A. Nelson[164], T.K. Nelson[144], S. Nemecek[127], P. Nemethy[110], A.A. Nepomuceno[24a], M. Nessi[30,aa], M.S. Neubauer[166],

M. Neumann[176], R.M. Neves[110], P. Nevski[25], P.R. Newman[18], D.H. Nguyen[6],
R.B. Nickerson[120], R. Nicolaidou[137], B. Nicquevert[30], J. Nielsen[138], N. Nikiforou[35],
A. Nikiforov[16], V. Nikolaenko[130,z], I. Nikolic-Audit[80], K. Nikolics[49], K. Nikolopoulos[18],
P. Nilsson[25], Y. Ninomiya[156], A. Nisati[133a], R. Nisius[101], T. Nobe[158], M. Nomachi[118],
I. Nomidis[29], S. Norberg[113], M. Nordberg[30], O. Novgorodova[44], S. Nowak[101], M. Nozaki[66],
L. Nozka[115], K. Ntekas[10], G. Nunes Hanninger[88], T. Nunnemann[100], E. Nurse[78],
F. Nuti[88], B.J. O'Brien[46], F. O'grady[7], D.C. O'Neil[143], V. O'Shea[53], F.G. Oakham[29,e],
H. Oberlack[101], T. Obermann[21], J. Ocariz[80], A. Ochi[67], I. Ochoa[78], S. Oda[70], S. Odaka[66],
H. Ogren[61], A. Oh[84], S.H. Oh[45], C.C. Ohm[15], H. Ohman[167], H. Oide[30], W. Okamura[118],
H. Okawa[161], Y. Okumura[31], T. Okuyama[156], A. Olariu[26a], A.G. Olchevski[65],
S.A. Olivares Pino[46], D. Oliveira Damazio[25], E. Oliver Garcia[168], A. Olszewski[39],
J. Olszowska[39], A. Onofre[126a,126e], P.U.E. Onyisi[31,o], C.J. Oram[160a], M.J. Oreglia[31],
Y. Oren[154], D. Orestano[135a,135b], N. Orlando[73a,73b], C. Oropeza Barrera[53], R.S. Orr[159],
B. Osculati[50a,50b], R. Ospanov[122], G. Otero y Garzon[27], H. Otono[70], M. Ouchrif[136d],
E.A. Ouellette[170], F. Ould-Saada[119], A. Ouraou[137], K.P. Oussoren[107], Q. Ouyang[33a],
A. Ovcharova[15], M. Owen[84], V.E. Ozcan[19a], N. Ozturk[8], K. Pachal[120],
A. Pacheco Pages[12], C. Padilla Aranda[12], M. Pagáčová[48], S. Pagan Griso[15], E. Paganis[140],
C. Pahl[101], F. Paige[25], P. Pais[86], K. Pajchel[119], G. Palacino[160b], S. Palestini[30],
M. Palka[38b], D. Pallin[34], A. Palma[126a,126b], J.D. Palmer[18], Y.B. Pan[174],
E. Panagiotopoulou[10], J.G. Panduro Vazquez[77], P. Pani[107], N. Panikashvili[89],
S. Panitkin[25], D. Pantea[26a], L. Paolozzi[134a,134b], Th.D. Papadopoulou[10],
K. Papageorgiou[155], A. Paramonov[6], D. Paredes Hernandez[155], M.A. Parker[28],
F. Parodi[50a,50b], J.A. Parsons[35], U. Parzefall[48], E. Pasqualucci[133a], S. Passaggio[50a],
A. Passeri[135a], F. Pastore[135a,135b,*], Fr. Pastore[77], G. Pásztor[29], S. Pataraia[176],
N.D. Patel[151], J.R. Pater[84], S. Patricelli[104a,104b], T. Pauly[30], J. Pearce[170],
L.E. Pedersen[36], M. Pedersen[119], S. Pedraza Lopez[168], R. Pedro[126a,126b],
S.V. Peleganchuk[109], D. Pelikan[167], H. Peng[33b], B. Penning[31], J. Penwell[61],
D.V. Perepelitsa[25], E. Perez Codina[160a], M.T. Pérez García-Estañ[168], L. Perini[91a,91b],
H. Pernegger[30], S. Perrella[104a,104b], R. Peschke[42], V.D. Peshekhonov[65], K. Peters[30],
R.F.Y. Peters[84], B.A. Petersen[30], T.C. Petersen[36], E. Petit[42], A. Petridis[147a,147b],
C. Petridou[155], E. Petrolo[133a], F. Petrucci[135a,135b], N.E. Pettersson[158], R. Pezoa[32b],
P.W. Phillips[131], G. Piacquadio[144], E. Pianori[171], A. Picazio[49], E. Piccaro[76],
M. Piccinini[20a,20b], M.A. Pickering[120], R. Piegaia[27], D.T. Pignotti[111], J.E. Pilcher[31],
A.D. Pilkington[78], J. Pina[126a,126b,126d], M. Pinamonti[165a,165c,ab], A. Pinder[120],
J.L. Pinfold[3], A. Pingel[36], B. Pinto[126a], S. Pires[80], M. Pitt[173], C. Pizio[91a,91b],
L. Plazak[145a], M.-A. Pleier[25], V. Pleskot[129], E. Plotnikova[65], P. Plucinski[147a,147b],
D. Pluth[64], S. Poddar[58a], F. Podlyski[34], R. Poettgen[83], L. Poggioli[117], D. Pohl[21],
M. Pohl[49], G. Polesello[121a], A. Policicchio[37a,37b], R. Polifka[159], A. Polini[20a],
C.S. Pollard[53], V. Polychronakos[25], K. Pommès[30], L. Pontecorvo[133a], B.G. Pope[90],
G.A. Popeneciu[26b], D.S. Popovic[13a], A. Poppleton[30], X. Portell Bueso[12], S. Pospisil[128],
K. Potamianos[15], I.N. Potrap[65], C.J. Potter[150], C.T. Potter[116], G. Poulard[30], J. Poveda[61],
V. Pozdnyakov[65], P. Pralavorio[85], A. Pranko[15], S. Prasad[30], R. Pravahan[8], S. Prell[64],
D. Price[84], J. Price[74], L.E. Price[6], D. Prieur[125], M. Primavera[73a], S. Prince[87],

M. Proissl[46], K. Prokofiev[47], F. Prokoshin[32b], E. Protopapadaki[137], S. Protopopescu[25], J. Proudfoot[6], M. Przybycien[38a], H. Przysiezniak[5], E. Ptacek[116], D. Puddu[135a,135b], E. Pueschel[86], D. Puldon[149], M. Purohit[25,ac], P. Puzo[117], J. Qian[89], G. Qin[53], Y. Qin[84], A. Quadt[54], D.R. Quarrie[15], W.B. Quayle[165a,165b], M. Queitsch-Maitland[84], D. Quilty[53], A. Qureshi[160b], V. Radeka[25], V. Radescu[42], S.K. Radhakrishnan[149], P. Radloff[116], P. Rados[88], F. Ragusa[91a,91b], G. Rahal[179], S. Rajagopalan[25], M. Rammensee[30], C. Rangel-Smith[167], K. Rao[164], F. Rauscher[100], T.C. Rave[48], T. Ravenscroft[53], M. Raymond[30], A.L. Read[119], N.P. Readioff[74], D.M. Rebuzzi[121a,121b], A. Redelbach[175], G. Redlinger[25], R. Reece[138], K. Reeves[41], L. Rehnisch[16], H. Reisin[27], M. Relich[164], C. Rembser[30], H. Ren[33a], Z.L. Ren[152], A. Renaud[117], M. Rescigno[133a], S. Resconi[91a], O.L. Rezanova[109,c], P. Reznicek[129], R. Rezvani[95], R. Richter[101], M. Ridel[80], P. Rieck[16], J. Rieger[54], M. Rijssenbeek[149], A. Rimoldi[121a,121b], L. Rinaldi[20a], E. Ritsch[62], I. Riu[12], F. Rizatdinova[114], E. Rizvi[76], S.H. Robertson[87,k], A. Robichaud-Veronneau[87], D. Robinson[28], J.E.M. Robinson[84], A. Robson[53], C. Roda[124a,124b], L. Rodrigues[30], S. Roe[30], O. Røhne[119], S. Rolli[162], A. Romaniouk[98], M. Romano[20a,20b], E. Romero Adam[168], N. Rompotis[139], M. Ronzani[48], L. Roos[80], E. Ros[168], S. Rosati[133a], K. Rosbach[49], M. Rose[77], P. Rose[138], P.L. Rosendahl[14], O. Rosenthal[142], V. Rossetti[147a,147b], E. Rossi[104a,104b], L.P. Rossi[50a], R. Rosten[139], M. Rotaru[26a], I. Roth[173], J. Rothberg[139], D. Rousseau[117], C.R. Royon[137], A. Rozanov[85], Y. Rozen[153], X. Ruan[146c], F. Rubbo[12], I. Rubinskiy[42], V.I. Rud[99], C. Rudolph[44], M.S. Rudolph[159], F. Rühr[48], A. Ruiz-Martinez[30], Z. Rurikova[48], N.A. Rusakovich[65], A. Ruschke[100], H.L. Russell[139], J.P. Rutherfoord[7], N. Ruthmann[48], Y.F. Ryabov[123], M. Rybar[129], G. Rybkin[117], N.C. Ryder[120], A.F. Saavedra[151], G. Sabato[107], S. Sacerdoti[27], A. Saddique[3], I. Sadeh[154], H.F-W. Sadrozinski[138], R. Sadykov[65], F. Safai Tehrani[133a], H. Sakamoto[156], Y. Sakurai[172], G. Salamanna[135a,135b], A. Salamon[134a], M. Saleem[113], D. Salek[107], P.H. Sales De Bruin[139], D. Salihagic[101], A. Salnikov[144], J. Salt[168], D. Salvatore[37a,37b], F. Salvatore[150], A. Salvucci[106], A. Salzburger[30], D. Sampsonidis[155], A. Sanchez[104a,104b], J. Sánchez[168], V. Sanchez Martinez[168], H. Sandaker[14], R.L. Sandbach[76], H.G. Sander[83], M.P. Sanders[100], M. Sandhoff[176], T. Sandoval[28], C. Sandoval[163], R. Sandstroem[101], D.P.C. Sankey[131], A. Sansoni[47], C. Santoni[34], R. Santonico[134a,134b], H. Santos[126a], I. Santoyo Castillo[150], K. Sapp[125], A. Sapronov[65], J.G. Saraiva[126a,126d], B. Sarrazin[21], G. Sartisohn[176], O. Sasaki[66], Y. Sasaki[156], G. Sauvage[5,*], E. Sauvan[5], P. Savard[159,e], D.O. Savu[30], C. Sawyer[120], L. Sawyer[79,m], D.H. Saxon[53], J. Saxon[31], C. Sbarra[20a], A. Sbrizzi[20a,20b], T. Scanlon[78], D.A. Scannicchio[164], M. Scarcella[151], V. Scarfone[37a,37b], J. Schaarschmidt[173], P. Schacht[101], D. Schaefer[30], R. Schaefer[42], S. Schaepe[21], S. Schaetzel[58b], U. Schäfer[83], A.C. Schaffer[117], D. Schaile[100], R.D. Schamberger[149], V. Scharf[58a], V.A. Schegelsky[123], D. Scheirich[129], M. Schernau[164], M.I. Scherzer[35], C. Schiavi[50a,50b], J. Schieck[100], C. Schillo[48], M. Schioppa[37a,37b], S. Schlenker[30], E. Schmidt[48], K. Schmieden[30], C. Schmitt[83], S. Schmitt[58b], B. Schneider[17], Y.J. Schnellbach[74], U. Schnoor[44], L. Schoeffel[137], A. Schoening[58b], B.D. Schoenrock[90], A.L.S. Schorlemmer[54], M. Schott[83], D. Schouten[160a], J. Schovancova[25], S. Schramm[159], M. Schreyer[175], C. Schroeder[83], N. Schuh[83], M.J. Schultens[21], H.-C. Schultz-Coulon[58a], H. Schulz[16], M. Schumacher[48],

B.A. Schumm[138], Ph. Schune[137], C. Schwanenberger[84], A. Schwartzman[144], T.A. Schwarz[89], Ph. Schwegler[101], Ph. Schwemling[137], R. Schwienhorst[90], J. Schwindling[137], T. Schwindt[21], M. Schwoerer[5], F.G. Sciacca[17], E. Scifo[117], G. Sciolla[23], F. Scuri[124a,124b], F. Scutti[21], J. Searcy[89], G. Sedov[42], E. Sedykh[123], P. Seema[21], S.C. Seidel[105], A. Seiden[138], F. Seifert[128], J.M. Seixas[24a], G. Sekhniaidze[104a], S.J. Sekula[40], K.E. Selbach[46], D.M. Seliverstov[123,*], G. Sellers[74], N. Semprini-Cesari[20a,20b], C. Serfon[30], L. Serin[117], L. Serkin[54], T. Serre[85], R. Seuster[160a], H. Severini[113], T. Sfiligoj[75], F. Sforza[101], A. Sfyrla[30], E. Shabalina[54], M. Shamim[116], L.Y. Shan[33a], R. Shang[166], J.T. Shank[22], M. Shapiro[15], P.B. Shatalov[97], K. Shaw[165a,165b], C.Y. Shehu[150], P. Sherwood[78], L. Shi[152,ad], S. Shimizu[67], C.O. Shimmin[164], M. Shimojima[102], M. Shiyakova[65], A. Shmeleva[96], D. Shoaleh Saadi[95], M.J. Shochet[31], D. Short[120], S. Shrestha[111], E. Shulga[98], M.A. Shupe[7], S. Shushkevich[42], P. Sicho[127], O. Sidiropoulou[155], D. Sidorov[114], A. Sidoti[133a], F. Siegert[44], Dj. Sijacki[13a], J. Silva[126a,126d], Y. Silver[154], D. Silverstein[144], S.B. Silverstein[147a], V. Simak[128], O. Simard[5], Lj. Simic[13a], S. Simion[117], E. Simioni[83], B. Simmons[78], D. Simon[34], R. Simoniello[91a,91b], P. Sinervo[159], N.B. Sinev[116], G. Siragusa[175], A. Sircar[79], A.N. Sisakyan[65,*], S.Yu. Sivoklokov[99], J. Sjölin[147a,147b], T.B. Sjursen[14], H.P. Skottowe[57], P. Skubic[113], M. Slater[18], T. Slavicek[128], M. Slawinska[107], K. Sliwa[162], V. Smakhtin[173], B.H. Smart[46], L. Smestad[14], S.Yu. Smirnov[98], Y. Smirnov[98], L.N. Smirnova[99,ae], O. Smirnova[81], K.M. Smith[53], M. Smizanska[72], K. Smolek[128], A.A. Snesarev[96], G. Snidero[76], S. Snyder[25], R. Sobie[170,k], F. Socher[44], A. Soffer[154], D.A. Soh[152,ad], C.A. Solans[30], M. Solar[128], J. Solc[128], E.Yu. Soldatov[98], U. Soldevila[168], A.A. Solodkov[130], A. Soloshenko[65], O.V. Solovyanov[130], V. Solovyev[123], P. Sommer[48], H.Y. Song[33b], N. Soni[1], A. Sood[15], A. Sopczak[128], B. Sopko[128], V. Sopko[128], V. Sorin[12], M. Sosebee[8], R. Soualah[165a,165c], P. Soueid[95], A.M. Soukharev[109,c], D. South[42], S. Spagnolo[73a,73b], F. Spanò[77], W.R. Spearman[57], F. Spettel[101], R. Spighi[20a], G. Spigo[30], L.A. Spiller[88], M. Spousta[129], T. Spreitzer[159], R.D. St. Denis[53,*], S. Staerz[44], J. Stahlman[122], R. Stamen[58a], S. Stamm[16], E. Stanecka[39], C. Stanescu[135a], M. Stanescu-Bellu[42], M.M. Stanitzki[42], S. Stapnes[119], E.A. Starchenko[130], J. Stark[55], P. Staroba[127], P. Starovoitov[42], R. Staszewski[39], P. Stavina[145a,*], P. Steinberg[25], B. Stelzer[143], H.J. Stelzer[30], O. Stelzer-Chilton[160a], H. Stenzel[52], S. Stern[101], G.A. Stewart[53], J.A. Stillings[21], M.C. Stockton[87], M. Stoebe[87], G. Stoicea[26a], P. Stolte[54], S. Stonjek[101], A.R. Stradling[8], A. Straessner[44], M.E. Stramaglia[17], J. Strandberg[148], S. Strandberg[147a,147b], A. Strandlie[119], E. Strauss[144], M. Strauss[113], P. Strizenec[145b], R. Ströhmer[175], D.M. Strom[116], R. Stroynowski[40], A. Strubig[106], S.A. Stucci[17], B. Stugu[14], N.A. Styles[42], D. Su[144], J. Su[125], R. Subramaniam[79], A. Succurro[12], Y. Sugaya[118], C. Suhr[108], M. Suk[128], V.V. Sulin[96], S. Sultansoy[4d], T. Sumida[68], S. Sun[57], X. Sun[33a], J.E. Sundermann[48], K. Suruliz[150], G. Susinno[37a,37b], M.R. Sutton[150], Y. Suzuki[66], M. Svatos[127], S. Swedish[169], M. Swiatlowski[144], I. Sykora[145a], T. Sykora[129], D. Ta[90], C. Taccini[135a,135b], K. Tackmann[42], J. Taenzer[159], A. Taffard[164], R. Tafirout[160a], N. Taiblum[154], H. Takai[25], R. Takashima[69], H. Takeda[67], T. Takeshita[141], Y. Takubo[66], M. Talby[85], A.A. Talyshev[109,c], J.Y.C. Tam[175], K.G. Tan[88], J. Tanaka[156], R. Tanaka[117], S. Tanaka[132], S. Tanaka[66], A.J. Tanasijczuk[143], B.B. Tannenwald[111], N. Tannoury[21],

S. Tapprogge[83], S. Tarem[153], F. Tarrade[29], G.F. Tartarelli[91a], P. Tas[129], M. Tasevsky[127], T. Tashiro[68], E. Tassi[37a,37b], A. Tavares Delgado[126a,126b], Y. Tayalati[136d], F.E. Taylor[94], G.N. Taylor[88], W. Taylor[160b], F.A. Teischinger[30], M. Teixeira Dias Castanheira[76], P. Teixeira-Dias[77], K.K. Temming[48], H. Ten Kate[30], P.K. Teng[152], J.J. Teoh[118], S. Terada[66], K. Terashi[156], J. Terron[82], S. Terzo[101], M. Testa[47], R.J. Teuscher[159,k], J. Therhaag[21], T. Theveneaux-Pelzer[34], J.P. Thomas[18], J. Thomas-Wilsker[77], E.N. Thompson[35], P.D. Thompson[18], R.J. Thompson[84], A.S. Thompson[53], L.A. Thomsen[36], E. Thomson[122], M. Thomson[28], W.M. Thong[88], R.P. Thun[89,*], F. Tian[35], M.J. Tibbetts[15], V.O. Tikhomirov[96,af], Yu.A. Tikhonov[109,c], S. Timoshenko[98], E. Tiouchichine[85], P. Tipton[177], S. Tisserant[85], T. Todorov[5], S. Todorova-Nova[129], J. Tojo[70], S. Tokár[145a], K. Tokushuku[66], K. Tollefson[90], E. Tolley[57], L. Tomlinson[84], M. Tomoto[103], L. Tompkins[31], K. Toms[105], N.D. Topilin[65], E. Torrence[116], H. Torres[143], E. Torró Pastor[168], J. Toth[85,ag], F. Touchard[85], D.R. Tovey[140], H.L. Tran[117], T. Trefzger[175], L. Tremblet[30], A. Tricoli[30], I.M. Trigger[160a], S. Trincaz-Duvoid[80], M.F. Tripiana[12], W. Trischuk[159], B. Trocmé[55], C. Troncon[91a], M. Trottier-McDonald[15], M. Trovatelli[135a,135b], P. True[90], M. Trzebinski[39], A. Trzupek[39], C. Tsarouchas[30], J.C-L. Tseng[120], P.V. Tsiareshka[92], D. Tsionou[137], G. Tsipolitis[10], N. Tsirintanis[9], S. Tsiskaridze[12], V. Tsiskaridze[48], E.G. Tskhadadze[51a], I.I. Tsukerman[97], V. Tsulaia[15], S. Tsuno[66], D. Tsybychev[149], A. Tudorache[26a], V. Tudorache[26a], A.N. Tuna[122], S.A. Tupputi[20a,20b], S. Turchikhin[99,ae], D. Turecek[128], I. Turk Cakir[4c], R. Turra[91a,91b], A.J. Turvey[40], P.M. Tuts[35], A. Tykhonov[49], M. Tylmad[147a,147b], M. Tyndel[131], K. Uchida[21], I. Ueda[156], R. Ueno[29], M. Ughetto[85], M. Ugland[14], M. Uhlenbrock[21], F. Ukegawa[161], G. Unal[30], A. Undrus[25], G. Unel[164], F.C. Ungaro[48], Y. Unno[66], C. Unverdorben[100], J. Urban[145b], D. Urbaniec[35], P. Urquijo[88], G. Usai[8], A. Usanova[62], L. Vacavant[85], V. Vacek[128], B. Vachon[87], N. Valencic[107], S. Valentinetti[20a,20b], A. Valero[168], L. Valery[34], S. Valkar[129], E. Valladolid Gallego[168], S. Vallecorsa[49], J.A. Valls Ferrer[168], W. Van Den Wollenberg[107], P.C. Van Der Deijl[107], R. van der Geer[107], H. van der Graaf[107], R. Van Der Leeuw[107], D. van der Ster[30], N. van Eldik[30], P. van Gemmeren[6], J. Van Nieuwkoop[143], I. van Vulpen[107], M.C. van Woerden[30], M. Vanadia[133a,133b], W. Vandelli[30], R. Vanguri[122], A. Vaniachine[6], P. Vankov[42], F. Vannucci[80], G. Vardanyan[178], R. Vari[133a], E.W. Varnes[7], T. Varol[86], D. Varouchas[80], A. Vartapetian[8], K.E. Varvell[151], F. Vazeille[34], T. Vazquez Schroeder[54], J. Veatch[7], F. Veloso[126a,126c], T. Velz[21], S. Veneziano[133a], A. Ventura[73a,73b], D. Ventura[86], M. Venturi[170], N. Venturi[159], A. Venturini[23], V. Vercesi[121a], M. Verducci[133a,133b], W. Verkerke[107], J.C. Vermeulen[107], A. Vest[44], M.C. Vetterli[143,e], O. Viazlo[81], I. Vichou[166], T. Vickey[146c,ah], O.E. Vickey Boeriu[146c], G.H.A. Viehhauser[120], S. Viel[169], R. Vigne[30], M. Villa[20a,20b], M. Villaplana Perez[91a,91b], E. Vilucchi[47], M.G. Vincter[29], V.B. Vinogradov[65], J. Virzi[15], I. Vivarelli[150], F. Vives Vaque[3], S. Vlachos[10], D. Vladoiu[100], M. Vlasak[128], A. Vogel[21], M. Vogel[32a], P. Vokac[128], G. Volpi[124a,124b], M. Volpi[88], H. von der Schmitt[101], H. von Radziewski[48], E. von Toerne[21], V. Vorobel[129], K. Vorobev[98], M. Vos[168], R. Voss[30], J.H. Vossebeld[74], N. Vranjes[137], M. Vranjes Milosavljevic[13a], V. Vrba[127], M. Vreeswijk[107], T. Vu Anh[48], R. Vuillermet[30], I. Vukotic[31], Z. Vykydal[128], P. Wagner[21], W. Wagner[176], H. Wahlberg[71], S. Wahrmund[44],

J. Wakabayashi[103], J. Walder[72], R. Walker[100], W. Walkowiak[142], R. Wall[177], P. Waller[74], B. Walsh[177], C. Wang[33c], C. Wang[45], F. Wang[174], H. Wang[15], H. Wang[40], J. Wang[42], J. Wang[33a], K. Wang[87], R. Wang[105], S.M. Wang[152], T. Wang[21], X. Wang[177], C. Wanotayaroj[116], A. Warburton[87], C.P. Ward[28], D.R. Wardrope[78], M. Warsinsky[48], A. Washbrook[46], C. Wasicki[42], P.M. Watkins[18], A.T. Watson[18], I.J. Watson[151], M.F. Watson[18], G. Watts[139], S. Watts[84], B.M. Waugh[78], S. Webb[84], M.S. Weber[17], S.W. Weber[175], J.S. Webster[31], A.R. Weidberg[120], B. Weinert[61], J. Weingarten[54], C. Weiser[48], H. Weits[107], P.S. Wells[30], T. Wenaus[25], D. Wendland[16], Z. Weng[152,ad], T. Wengler[30], S. Wenig[30], N. Wermes[21], M. Werner[48], P. Werner[30], M. Wessels[58a], J. Wetter[162], K. Whalen[29], A. White[8], M.J. White[1], R. White[32b], S. White[124a,124b], D. Whiteson[164], D. Wicke[176], F.J. Wickens[131], W. Wiedenmann[174], M. Wielers[131], P. Wienemann[21], C. Wiglesworth[36], L.A.M. Wiik-Fuchs[21], P.A. Wijeratne[78], A. Wildauer[101], M.A. Wildt[42,ai], H.G. Wilkens[30], H.H. Williams[122], S. Williams[28], C. Willis[90], S. Willocq[86], A. Wilson[89], J.A. Wilson[18], I. Wingerter-Seez[5], F. Winklmeier[116], B.T. Winter[21], M. Wittgen[144], T. Wittig[43], J. Wittkowski[100], S.J. Wollstadt[83], M.W. Wolter[39], H. Wolters[126a,126c], B.K. Wosiek[39], J. Wotschack[30], M.J. Woudstra[84], K.W. Wozniak[39], M. Wright[53], M. Wu[55], S.L. Wu[174], X. Wu[49], Y. Wu[89], E. Wulf[35], T.R. Wyatt[84], B.M. Wynne[46], S. Xella[36], M. Xiao[137], D. Xu[33a], L. Xu[33b,aj], B. Yabsley[151], S. Yacoob[146b,ak], R. Yakabe[67], M. Yamada[66], H. Yamaguchi[156], Y. Yamaguchi[118], A. Yamamoto[66], S. Yamamoto[156], T. Yamamura[156], T. Yamanaka[156], K. Yamauchi[103], Y. Yamazaki[67], Z. Yan[22], H. Yang[33e], H. Yang[174], Y. Yang[111], S. Yanush[93], L. Yao[33a], W-M. Yao[15], Y. Yasu[66], E. Yatsenko[42], K.H. Yau Wong[21], J. Ye[40], S. Ye[25], I. Yeletskikh[65], A.L. Yen[57], E. Yildirim[42], M. Yilmaz[4b], R. Yoosoofmiya[125], K. Yorita[172], R. Yoshida[6], K. Yoshihara[156], C. Young[144], C.J.S. Young[30], S. Youssef[22], D.R. Yu[15], J. Yu[8], J.M. Yu[89], J. Yu[114], L. Yuan[67], A. Yurkewicz[108], I. Yusuff[28,al], B. Zabinski[39], R. Zaidan[63], A.M. Zaitsev[130,z], A. Zaman[149], S. Zambito[23], L. Zanello[133a,133b], D. Zanzi[88], C. Zeitnitz[176], M. Zeman[128], A. Zemla[38a], K. Zengel[23], O. Zenin[130], T. Ženiš[145a], D. Zerwas[117], G. Zevi della Porta[57], D. Zhang[89], F. Zhang[174], H. Zhang[90], J. Zhang[6], L. Zhang[152], R. Zhang[33b], X. Zhang[33d], Z. Zhang[117], Y. Zhao[33d], Z. Zhao[33b], A. Zhemchugov[65], J. Zhong[120], B. Zhou[89], L. Zhou[35], L. Zhou[40], N. Zhou[164], C.G. Zhu[33d], H. Zhu[33a], J. Zhu[89], Y. Zhu[33b], X. Zhuang[33a], K. Zhukov[96], A. Zibell[175], D. Zieminska[61], N.I. Zimine[65], C. Zimmermann[83], R. Zimmermann[21], S. Zimmermann[21], S. Zimmermann[48], Z. Zinonos[54], M. Ziolkowski[142], G. Zobernig[174], A. Zoccoli[20a,20b], M. zur Nedden[16], G. Zurzolo[104a,104b], L. Zwalinski[30].

[1] Department of Physics, University of Adelaide, Adelaide, Australia

[2] Physics Department, SUNY Albany, Albany NY, United States of America

[3] Department of Physics, University of Alberta, Edmonton AB, Canada

[4] [(a)] Department of Physics, Ankara University, Ankara; [(b)] Department of Physics, Gazi University, Ankara; [(c)] Istanbul Aydin University, Istanbul; [(d)] Division of Physics, TOBB University of Economics and Technology, Ankara, Turkey

[5] LAPP, CNRS/IN2P3 and Université de Savoie, Annecy-le-Vieux, France

[6] High Energy Physics Division, Argonne National Laboratory, Argonne IL, United States

of America
[7] Department of Physics, University of Arizona, Tucson AZ, United States of America
[8] Department of Physics, The University of Texas at Arlington, Arlington TX, United States of America
[9] Physics Department, University of Athens, Athens, Greece
[10] Physics Department, National Technical University of Athens, Zografou, Greece
[11] Institute of Physics, Azerbaijan Academy of Sciences, Baku, Azerbaijan
[12] Institut de Física d'Altes Energies and Departament de Física de la Universitat Autònoma de Barcelona, Barcelona, Spain
[13] [(a)] Institute of Physics, University of Belgrade, Belgrade; [(b)] Vinca Institute of Nuclear Sciences, University of Belgrade, Belgrade, Serbia
[14] Department for Physics and Technology, University of Bergen, Bergen, Norway
[15] Physics Division, Lawrence Berkeley National Laboratory and University of California, Berkeley CA, United States of America
[16] Department of Physics, Humboldt University, Berlin, Germany
[17] Albert Einstein Center for Fundamental Physics and Laboratory for High Energy Physics, University of Bern, Bern, Switzerland
[18] School of Physics and Astronomy, University of Birmingham, Birmingham, United Kingdom
[19] [(a)] Department of Physics, Bogazici University, Istanbul; [(b)] Department of Physics, Dogus University, Istanbul; [(c)] Department of Physics Engineering, Gaziantep University, Gaziantep, Turkey
[20] [(a)] INFN Sezione di Bologna; [(b)] Dipartimento di Fisica e Astronomia, Università di Bologna, Bologna, Italy
[21] Physikalisches Institut, University of Bonn, Bonn, Germany
[22] Department of Physics, Boston University, Boston MA, United States of America
[23] Department of Physics, Brandeis University, Waltham MA, United States of America
[24] [(a)] Universidade Federal do Rio De Janeiro COPPE/EE/IF, Rio de Janeiro; [(b)] Electrical Circuits Department, Federal University of Juiz de Fora (UFJF), Juiz de Fora; [(c)] Federal University of Sao Joao del Rei (UFSJ), Sao Joao del Rei; [(d)] Instituto de Fisica, Universidade de Sao Paulo, Sao Paulo, Brazil
[25] Physics Department, Brookhaven National Laboratory, Upton NY, United States of America
[26] [(a)] National Institute of Physics and Nuclear Engineering, Bucharest; [(b)] National Institute for Research and Development of Isotopic and Molecular Technologies, Physics Department, Cluj Napoca; [(c)] University Politehnica Bucharest, Bucharest; [(d)] West University in Timisoara, Timisoara, Romania
[27] Departamento de Física, Universidad de Buenos Aires, Buenos Aires, Argentina
[28] Cavendish Laboratory, University of Cambridge, Cambridge, United Kingdom
[29] Department of Physics, Carleton University, Ottawa ON, Canada
[30] CERN, Geneva, Switzerland
[31] Enrico Fermi Institute, University of Chicago, Chicago IL, United States of America
[32] [(a)] Departamento de Física, Pontificia Universidad Católica de Chile, Santiago; [(b)]

Departamento de Física, Universidad Técnica Federico Santa María, Valparaíso, Chile
[33] [(a)] Institute of High Energy Physics, Chinese Academy of Sciences, Beijing; [(b)] Department of Modern Physics, University of Science and Technology of China, Anhui; [(c)] Department of Physics, Nanjing University, Jiangsu; [(d)] School of Physics, Shandong University, Shandong; [(e)] Physics Department, Shanghai Jiao Tong University, Shanghai; [(f)] Physics Department, Tsinghua University, Beijing 100084, China
[34] Laboratoire de Physique Corpusculaire, Clermont Université and Université Blaise Pascal and CNRS/IN2P3, Clermont-Ferrand, France
[35] Nevis Laboratory, Columbia University, Irvington NY, United States of America
[36] Niels Bohr Institute, University of Copenhagen, Kobenhavn, Denmark
[37] [(a)] INFN Gruppo Collegato di Cosenza, Laboratori Nazionali di Frascati; [(b)] Dipartimento di Fisica, Università della Calabria, Rende, Italy
[38] [(a)] AGH University of Science and Technology, Faculty of Physics and Applied Computer Science, Krakow; [(b)] Marian Smoluchowski Institute of Physics, Jagiellonian University, Krakow, Poland
[39] The Henryk Niewodniczanski Institute of Nuclear Physics, Polish Academy of Sciences, Krakow, Poland
[40] Physics Department, Southern Methodist University, Dallas TX, United States of America
[41] Physics Department, University of Texas at Dallas, Richardson TX, United States of America
[42] DESY, Hamburg and Zeuthen, Germany
[43] Institut für Experimentelle Physik IV, Technische Universität Dortmund, Dortmund, Germany
[44] Institut für Kern- und Teilchenphysik, Technische Universität Dresden, Dresden, Germany
[45] Department of Physics, Duke University, Durham NC, United States of America
[46] SUPA - School of Physics and Astronomy, University of Edinburgh, Edinburgh, United Kingdom
[47] INFN Laboratori Nazionali di Frascati, Frascati, Italy
[48] Fakultät für Mathematik und Physik, Albert-Ludwigs-Universität, Freiburg, Germany
[49] Section de Physique, Université de Genève, Geneva, Switzerland
[50] [(a)] INFN Sezione di Genova; [(b)] Dipartimento di Fisica, Università di Genova, Genova, Italy
[51] [(a)] E. Andronikashvili Institute of Physics, Iv. Javakhishvili Tbilisi State University, Tbilisi; [(b)] High Energy Physics Institute, Tbilisi State University, Tbilisi, Georgia
[52] II Physikalisches Institut, Justus-Liebig-Universität Giessen, Giessen, Germany
[53] SUPA - School of Physics and Astronomy, University of Glasgow, Glasgow, United Kingdom
[54] II Physikalisches Institut, Georg-August-Universität, Göttingen, Germany
[55] Laboratoire de Physique Subatomique et de Cosmologie, Université Grenoble-Alpes, CNRS/IN2P3, Grenoble, France
[56] Department of Physics, Hampton University, Hampton VA, United States of America

[57] Laboratory for Particle Physics and Cosmology, Harvard University, Cambridge MA, United States of America
[58] [(a)] Kirchhoff-Institut für Physik, Ruprecht-Karls-Universität Heidelberg, Heidelberg; [(b)] Physikalisches Institut, Ruprecht-Karls-Universität Heidelberg, Heidelberg; [(c)] ZITI Institut für technische Informatik, Ruprecht-Karls-Universität Heidelberg, Mannheim, Germany
[59] Faculty of Applied Information Science, Hiroshima Institute of Technology, Hiroshima, Japan
[60] [(a)] Department of Physics, The Chinese University of Hong Kong, Shatin, N.T., Hong Kong; [(b)] Department of Physics, The University of Hong Kong, Hong Kong; [(c)] Department of Physics, The Hong Kong University of Science and Technology, Clear Water Bay, Kowloon, Hong Kong, China
[61] Department of Physics, Indiana University, Bloomington IN, United States of America
[62] Institut für Astro- und Teilchenphysik, Leopold-Franzens-Universität, Innsbruck, Austria
[63] University of Iowa, Iowa City IA, United States of America
[64] Department of Physics and Astronomy, Iowa State University, Ames IA, United States of America
[65] Joint Institute for Nuclear Research, JINR Dubna, Dubna, Russia
[66] KEK, High Energy Accelerator Research Organization, Tsukuba, Japan
[67] Graduate School of Science, Kobe University, Kobe, Japan
[68] Faculty of Science, Kyoto University, Kyoto, Japan
[69] Kyoto University of Education, Kyoto, Japan
[70] Department of Physics, Kyushu University, Fukuoka, Japan
[71] Instituto de Física La Plata, Universidad Nacional de La Plata and CONICET, La Plata, Argentina
[72] Physics Department, Lancaster University, Lancaster, United Kingdom
[73] [(a)] INFN Sezione di Lecce; [(b)] Dipartimento di Matematica e Fisica, Università del Salento, Lecce, Italy
[74] Oliver Lodge Laboratory, University of Liverpool, Liverpool, United Kingdom
[75] Department of Physics, Jožef Stefan Institute and University of Ljubljana, Ljubljana, Slovenia
[76] School of Physics and Astronomy, Queen Mary University of London, London, United Kingdom
[77] Department of Physics, Royal Holloway University of London, Surrey, United Kingdom
[78] Department of Physics and Astronomy, University College London, London, United Kingdom
[79] Louisiana Tech University, Ruston LA, United States of America
[80] Laboratoire de Physique Nucléaire et de Hautes Energies, UPMC and Université Paris-Diderot and CNRS/IN2P3, Paris, France
[81] Fysiska institutionen, Lunds universitet, Lund, Sweden
[82] Departamento de Fisica Teorica C-15, Universidad Autonoma de Madrid, Madrid, Spain

[83] Institut für Physik, Universität Mainz, Mainz, Germany
[84] School of Physics and Astronomy, University of Manchester, Manchester, United Kingdom
[85] CPPM, Aix-Marseille Université and CNRS/IN2P3, Marseille, France
[86] Department of Physics, University of Massachusetts, Amherst MA, United States of America
[87] Department of Physics, McGill University, Montreal QC, Canada
[88] School of Physics, University of Melbourne, Victoria, Australia
[89] Department of Physics, The University of Michigan, Ann Arbor MI, United States of America
[90] Department of Physics and Astronomy, Michigan State University, East Lansing MI, United States of America
[91] [(a)] INFN Sezione di Milano; [(b)] Dipartimento di Fisica, Università di Milano, Milano, Italy
[92] B.I. Stepanov Institute of Physics, National Academy of Sciences of Belarus, Minsk, Republic of Belarus
[93] National Scientific and Educational Centre for Particle and High Energy Physics, Minsk, Republic of Belarus
[94] Department of Physics, Massachusetts Institute of Technology, Cambridge MA, United States of America
[95] Group of Particle Physics, University of Montreal, Montreal QC, Canada
[96] P.N. Lebedev Institute of Physics, Academy of Sciences, Moscow, Russia
[97] Institute for Theoretical and Experimental Physics (ITEP), Moscow, Russia
[98] National Research Nuclear University MEPhI, Moscow, Russia
[99] D.V.Skobeltsyn Institute of Nuclear Physics, M.V.Lomonosov Moscow State University, Moscow, Russia
[100] Fakultät für Physik, Ludwig-Maximilians-Universität München, München, Germany
[101] Max-Planck-Institut für Physik (Werner-Heisenberg-Institut), München, Germany
[102] Nagasaki Institute of Applied Science, Nagasaki, Japan
[103] Graduate School of Science and Kobayashi-Maskawa Institute, Nagoya University, Nagoya, Japan
[104] [(a)] INFN Sezione di Napoli; [(b)] Dipartimento di Fisica, Università di Napoli, Napoli, Italy
[105] Department of Physics and Astronomy, University of New Mexico, Albuquerque NM, United States of America
[106] Institute for Mathematics, Astrophysics and Particle Physics, Radboud University Nijmegen/Nikhef, Nijmegen, Netherlands
[107] Nikhef National Institute for Subatomic Physics and University of Amsterdam, Amsterdam, Netherlands
[108] Department of Physics, Northern Illinois University, DeKalb IL, United States of America
[109] Budker Institute of Nuclear Physics, SB RAS, Novosibirsk, Russia
[110] Department of Physics, New York University, New York NY, United States of America

[111] Ohio State University, Columbus OH, United States of America
[112] Faculty of Science, Okayama University, Okayama, Japan
[113] Homer L. Dodge Department of Physics and Astronomy, University of Oklahoma, Norman OK, United States of America
[114] Department of Physics, Oklahoma State University, Stillwater OK, United States of America
[115] Palacký University, RCPTM, Olomouc, Czech Republic
[116] Center for High Energy Physics, University of Oregon, Eugene OR, United States of America
[117] LAL, Université Paris-Sud and CNRS/IN2P3, Orsay, France
[118] Graduate School of Science, Osaka University, Osaka, Japan
[119] Department of Physics, University of Oslo, Oslo, Norway
[120] Department of Physics, Oxford University, Oxford, United Kingdom
[121] [(a)] INFN Sezione di Pavia; [(b)] Dipartimento di Fisica, Università di Pavia, Pavia, Italy
[122] Department of Physics, University of Pennsylvania, Philadelphia PA, United States of America
[123] Petersburg Nuclear Physics Institute, Gatchina, Russia
[124] [(a)] INFN Sezione di Pisa; [(b)] Dipartimento di Fisica E. Fermi, Università di Pisa, Pisa, Italy
[125] Department of Physics and Astronomy, University of Pittsburgh, Pittsburgh PA, United States of America
[126] [(a)] Laboratorio de Instrumentacao e Fisica Experimental de Particulas - LIP, Lisboa; [(b)] Faculdade de Ciências, Universidade de Lisboa, Lisboa; [(c)] Department of Physics, University of Coimbra, Coimbra; [(d)] Centro de Física Nuclear da Universidade de Lisboa, Lisboa; [(e)] Departamento de Fisica, Universidade do Minho, Braga; [(f)] Departamento de Fisica Teorica y del Cosmos and CAFPE, Universidad de Granada, Granada (Spain); [(g)] Dep Fisica and CEFITEC of Faculdade de Ciencias e Tecnologia, Universidade Nova de Lisboa, Caparica, Portugal
[127] Institute of Physics, Academy of Sciences of the Czech Republic, Praha, Czech Republic
[128] Czech Technical University in Prague, Praha, Czech Republic
[129] Faculty of Mathematics and Physics, Charles University in Prague, Praha, Czech Republic
[130] State Research Center Institute for High Energy Physics, Protvino, Russia
[131] Particle Physics Department, Rutherford Appleton Laboratory, Didcot, United Kingdom
[132] Ritsumeikan University, Kusatsu, Shiga, Japan
[133] [(a)] INFN Sezione di Roma; [(b)] Dipartimento di Fisica, Sapienza Università di Roma, Roma, Italy
[134] [(a)] INFN Sezione di Roma Tor Vergata; [(b)] Dipartimento di Fisica, Università di Roma Tor Vergata, Roma, Italy
[135] [(a)] INFN Sezione di Roma Tre; [(b)] Dipartimento di Matematica e Fisica, Università Roma Tre, Roma, Italy

[136] [(a)] Faculté des Sciences Ain Chock, Réseau Universitaire de Physique des Hautes Energies - Université Hassan II, Casablanca; [(b)] Centre National de l'Energie des Sciences Techniques Nucleaires, Rabat; [(c)] Faculté des Sciences Semlalia, Université Cadi Ayyad, LPHEA-Marrakech; [(d)] Faculté des Sciences, Université Mohamed Premier and LPTPM, Oujda; [(e)] Faculté des sciences, Université Mohammed V-Agdal, Rabat, Morocco
[137] DSM/IRFU (Institut de Recherches sur les Lois Fondamentales de l'Univers), CEA Saclay (Commissariat à l'Energie Atomique et aux Energies Alternatives), Gif-sur-Yvette, France
[138] Santa Cruz Institute for Particle Physics, University of California Santa Cruz, Santa Cruz CA, United States of America
[139] Department of Physics, University of Washington, Seattle WA, United States of America
[140] Department of Physics and Astronomy, University of Sheffield, Sheffield, United Kingdom
[141] Department of Physics, Shinshu University, Nagano, Japan
[142] Fachbereich Physik, Universität Siegen, Siegen, Germany
[143] Department of Physics, Simon Fraser University, Burnaby BC, Canada
[144] SLAC National Accelerator Laboratory, Stanford CA, United States of America
[145] [(a)] Faculty of Mathematics, Physics & Informatics, Comenius University, Bratislava; [(b)] Department of Subnuclear Physics, Institute of Experimental Physics of the Slovak Academy of Sciences, Kosice, Slovak Republic
[146] [(a)] Department of Physics, University of Cape Town, Cape Town; [(b)] Department of Physics, University of Johannesburg, Johannesburg; [(c)] School of Physics, University of the Witwatersrand, Johannesburg, South Africa
[147] [(a)] Department of Physics, Stockholm University; [(b)] The Oskar Klein Centre, Stockholm, Sweden
[148] Physics Department, Royal Institute of Technology, Stockholm, Sweden
[149] Departments of Physics & Astronomy and Chemistry, Stony Brook University, Stony Brook NY, United States of America
[150] Department of Physics and Astronomy, University of Sussex, Brighton, United Kingdom
[151] School of Physics, University of Sydney, Sydney, Australia
[152] Institute of Physics, Academia Sinica, Taipei, Taiwan
[153] Department of Physics, Technion: Israel Institute of Technology, Haifa, Israel
[154] Raymond and Beverly Sackler School of Physics and Astronomy, Tel Aviv University, Tel Aviv, Israel
[155] Department of Physics, Aristotle University of Thessaloniki, Thessaloniki, Greece
[156] International Center for Elementary Particle Physics and Department of Physics, The University of Tokyo, Tokyo, Japan
[157] Graduate School of Science and Technology, Tokyo Metropolitan University, Tokyo, Japan
[158] Department of Physics, Tokyo Institute of Technology, Tokyo, Japan
[159] Department of Physics, University of Toronto, Toronto ON, Canada

[160] [(a)] TRIUMF, Vancouver BC; [(b)] Department of Physics and Astronomy, York University, Toronto ON, Canada
[161] Faculty of Pure and Applied Sciences, University of Tsukuba, Tsukuba, Japan
[162] Department of Physics and Astronomy, Tufts University, Medford MA, United States of America
[163] Centro de Investigaciones, Universidad Antonio Narino, Bogota, Colombia
[164] Department of Physics and Astronomy, University of California Irvine, Irvine CA, United States of America
[165] [(a)] INFN Gruppo Collegato di Udine, Sezione di Trieste, Udine; [(b)] ICTP, Trieste; [(c)] Dipartimento di Chimica, Fisica e Ambiente, Università di Udine, Udine, Italy
[166] Department of Physics, University of Illinois, Urbana IL, United States of America
[167] Department of Physics and Astronomy, University of Uppsala, Uppsala, Sweden
[168] Instituto de Física Corpuscular (IFIC) and Departamento de Física Atómica, Molecular y Nuclear and Departamento de Ingeniería Electrónica and Instituto de Microelectrónica de Barcelona (IMB-CNM), University of Valencia and CSIC, Valencia, Spain
[169] Department of Physics, University of British Columbia, Vancouver BC, Canada
[170] Department of Physics and Astronomy, University of Victoria, Victoria BC, Canada
[171] Department of Physics, University of Warwick, Coventry, United Kingdom
[172] Waseda University, Tokyo, Japan
[173] Department of Particle Physics, The Weizmann Institute of Science, Rehovot, Israel
[174] Department of Physics, University of Wisconsin, Madison WI, United States of America
[175] Fakultät für Physik und Astronomie, Julius-Maximilians-Universität, Würzburg, Germany
[176] Fachbereich C Physik, Bergische Universität Wuppertal, Wuppertal, Germany
[177] Department of Physics, Yale University, New Haven CT, United States of America
[178] Yerevan Physics Institute, Yerevan, Armenia
[179] Centre de Calcul de l'Institut National de Physique Nucléaire et de Physique des Particules (IN2P3), Villeurbanne, France
[a] Also at Department of Physics, King's College London, London, United Kingdom
[b] Also at Institute of Physics, Azerbaijan Academy of Sciences, Baku, Azerbaijan
[c] Also at Novosibirsk State University, Novosibirsk, Russia
[d] Also at Particle Physics Department, Rutherford Appleton Laboratory, Didcot, United Kingdom
[e] Also at TRIUMF, Vancouver BC, Canada
[f] Also at Department of Physics, California State University, Fresno CA, United States of America
[g] Also at Department of Physics, University of Fribourg, Fribourg, Switzerland
[h] Also at Tomsk State University, Tomsk, Russia
[i] Also at CPPM, Aix-Marseille Université and CNRS/IN2P3, Marseille, France
[j] Also at Università di Napoli Parthenope, Napoli, Italy
[k] Also at Institute of Particle Physics (IPP), Canada

[l] Also at Department of Physics, St. Petersburg State Polytechnical University, St. Petersburg, Russia
[m] Also at Louisiana Tech University, Ruston LA, United States of America
[n] Also at Institucio Catalana de Recerca i Estudis Avancats, ICREA, Barcelona, Spain
[o] Also at Department of Physics, The University of Texas at Austin, Austin TX, United States of America
[p] Also at Institute of Theoretical Physics, Ilia State University, Tbilisi, Georgia
[q] Also at CERN, Geneva, Switzerland
[r] Also at Ochadai Academic Production, Ochanomizu University, Tokyo, Japan
[s] Also at Manhattan College, New York NY, United States of America
[t] Also at Institute of Physics, Academia Sinica, Taipei, Taiwan
[u] Also at LAL, Université Paris-Sud and CNRS/IN2P3, Orsay, France
[v] Also at Academia Sinica Grid Computing, Institute of Physics, Academia Sinica, Taipei, Taiwan
[w] Also at Laboratoire de Physique Nucléaire et de Hautes Energies, UPMC and Université Paris-Diderot and CNRS/IN2P3, Paris, France
[x] Also at School of Physical Sciences, National Institute of Science Education and Research, Bhubaneswar, India
[y] Also at Dipartimento di Fisica, Sapienza Università di Roma, Roma, Italy
[z] Also at Moscow Institute of Physics and Technology State University, Dolgoprudny, Russia
[aa] Also at Section de Physique, Université de Genève, Geneva, Switzerland
[ab] Also at International School for Advanced Studies (SISSA), Trieste, Italy
[ac] Also at Department of Physics and Astronomy, University of South Carolina, Columbia SC, United States of America
[ad] Also at School of Physics and Engineering, Sun Yat-sen University, Guangzhou, China
[ae] Also at Faculty of Physics, M.V.Lomonosov Moscow State University, Moscow, Russia
[af] Also at National Research Nuclear University MEPhI, Moscow, Russia
[ag] Also at Institute for Particle and Nuclear Physics, Wigner Research Centre for Physics, Budapest, Hungary
[ah] Also at Department of Physics, Oxford University, Oxford, United Kingdom
[ai] Also at Institut für Experimentalphysik, Universität Hamburg, Hamburg, Germany
[aj] Also at Department of Physics, The University of Michigan, Ann Arbor MI, United States of America
[ak] Also at Discipline of Physics, University of KwaZulu-Natal, Durban, South Africa
[al] Also at University of Malaya, Department of Physics, Kuala Lumpur, Malaysia
* Deceased